\documentclass[a4paper,10pt]{article}
\pdfoutput=1
\usepackage[utf8x]{inputenc}
\usepackage{float}
\usepackage{amsmath}
\usepackage{fullpage}
\usepackage{titlesec}
\usepackage{color}
\usepackage{bbm}
\usepackage{listings}
\usepackage{amssymb}
\usepackage{epsfig}
\usepackage{mathtools}
\usepackage{slashed}
\usepackage{breqn}
\usepackage{graphicx}
\usepackage{lscape}
\usepackage{longtable}
\usepackage[colorlinks,citecolor=red,linkcolor=blue]{hyperref}
\usepackage{hvfloat}

\titleformat{\paragraph}
{\normalfont\normalsize\bfseries}{\theparagraph}{1em}{}
\titlespacing*{\paragraph}
{0pt}{3.25ex plus 1ex minus .2ex}{1.5ex plus .2ex}

\renewcommand{\thesection}{\arabic{section}}

\def\beq{\begin{equation}}
\def\eeq{\end{equation}}
\def\bea{\begin{eqnarray}}
\def\eea{\end{eqnarray}}
\def\bmat{\begin{pmatrix}}
\def\emat{\end{pmatrix}}

\def\to{\rightarrow}

\title{
Status and Prospects of the Two-Higgs-Doublet SU(6)/Sp(6) Little-Higgs Model 
and the Alignment Limit
}

\author{Shrihari~Gopalakrishna~\thanks{shri@imsc.res.in}~, 
Tuhin Subhra Mukherjee~\thanks{tuhin@imsc.res.in}~, 
Soumya Sadhukhan~\thanks{soumyasad@imsc.res.in}~, \\  
\small{The Institute of Mathematical Sciences (IMSc),} \\ 
\small{C.I.T Campus, Taramani, Chennai 600113, India.}
}

\begin{document}
\maketitle

\begin{abstract}

We study in detail the little-Higgs model proposed by Low, Skiba and Smith with an SU(6)/Sp(6) group structure. 
The effective theory at the TeV scale is a two-Higgs doublet model (2HDM) with additional heavy vector-like fermions and vector-bosons.  
We identify a set of independent input parameters and develop expressions for masses and couplings in terms of these.
We perform a random scan of the parameter space and find points 
that satisfy constraints, including the recent 8~TeV LHC Higgs measurements, namely, 
the Higgs mass, Higgs couplings to the top, bottom, $\tau$, $W^\pm$ and $Z$,
top-quark mass, and collider bounds on colored vector-like fermions ($t'$ and $b'$), 
and also precision electroweak constraints.
The LHC constraints on the $hWW$ and $hZZ$ couplings are satisfied by being close to the ``alignment limit''. 
We find how fine-tuned the model is after including these constraints. 
For the points that satisfy the constraints, we present the 1-loop effective couplings of the CP-even and CP-odd neutral scalars to two gluons 
including contributions of standard model and heavy vector-like quarks. 
We also present 
the branching ratios of the heavy neutral scalars into the $\gamma\gamma,\, \tau\bar\tau,\, b\bar b,\, t\bar t, WW, ZZ, Zh, hh$ modes,
and the heavy charged scalar into $tb,\, \tau\nu, cs, W h$ modes. 
These will aid searches of the heavy scalars at the LHC and other future colliders. 
 
\end{abstract}

\section{Introduction}
The little-Higgs model~\cite{ArkaniHamed:2001nc} is an attractive possibility to stabilize the electroweak scale 
by preventing the Higgs mass from receiving quadratically divergent corrections at the 1-loop level.
This is realized by making the Higgs a pseudo-Nambu-Goldstone boson (pNGB) (for reviews of many models with this idea, see Refs.~\cite{Schmaltz:2005ky,Perelstein:2005ka}). 
A global symmetry ($\cal{G}$), containing the standard model (SM) gauge group is imposed on the Lagrangian,
 which is spontaneously broken to a subgroup $\cal{H}$
giving rise to Nambu-Goldstone bosons (NGB) which are massless at the tree-level and live in the coset $\cal{G}/\cal{H}$.
The Higgs boson is one such NGB in the little-Higgs framework. 
The gauge and Yukawa interactions explicitly break some of these global symmetries,
due to which the scalars including the Higgs picks up a mass at the loop-level, making it a pNGB.
The breaking is specially arranged so that the mass it picks up is finite at 1-loop. 

To implement this mechanism, the scalar multiplet that contains the Higgs fills out a representation of the global symmetry group,
which therefore contain additional scalars.
For the same reason, new fermion states beyond the SM are also introduced to have enough symmetries to prevent quadratic divergence at 1-loop.
The beyond the standard model (BSM) fermions are made heavy by making them vector-like with respect to the SM gauge group. 
Typically, under the SM $SU(2)$, the extra scalar states are singlets, doublets, or triplets, while the vector-like fermions are singlets or doublets.
Depending on the $\cal{G}$ and $\cal{H}$ and the way the global symmetries are broken, these extra scalar states could be ``light'' like the Higgs (with $m_h \ll f$), 
or could be heavy with mass around $f$.  
Precision electroweak constraints impose non-trivial constraints on little-Higgs models, somewhat fine-tuning the models. 
Imposing the $T$-parity has been shown~\cite{Cheng:2003ju} to alleviate this problem. 

Well-studied little-Higgs models include the ``minimal-moose''~\cite{ArkaniHamed:2002qx} and the ``littlest-Higgs''~\cite{ArkaniHamed:2002qy}.
Many of these models contain a two-Higgs-doublet model (2HDM) structure and it is interesting to ask how these fare given the most recent LHC 8~TeV data.
In Ref.~\cite{Gopalakrishna:2015wwa} we discussed this question in a model-independent setting, while also including some effective vector-like fermion models.
This paper focuses on the little-Higgs model with a 2HDM structure.
Some little-Higgs models that contain a 2HDM structure are:
\begin{itemize}

\item[]
{The minimal-moose with T-parity by Cheng and Low (Ref.~\cite{Cheng:2004yc})}:
Unlike in the original minimal-moose where $SU(3)\times SU(2)\times U(1)$ was gauged,
in this model, to implement $T$-parity, $[SU(2)\times U(1)]^2$ is gauged, and the diagonal sub-group is identified with the SM gauge-group. 
The low-energy effective theory much below the scale $f$ is a 2HDM. 
In the fermionic sector, this model contains a new $SU(2)$ singlet vector-like Weyl-fermion pair ($u^\prime$, ${u^c}^\prime$), i.e. one additional singlet Dirac fermion
(whose mass is given by $\lambda^\prime f$).

\item[]
{The minimal-moose with T-parity by Cheng and Low (Ref.~\cite{Cheng:2003ju})}:
This model can be thought of as a UV completion of the above model in Ref.~\cite{Cheng:2004yc}.
The $SU(3)$ global symmetry structure in the minimal moose is enlarged to $SO(5)$ in order to 
include the custodial symmetry group $SU(2)_c$ which further keeps the T-parameter under control. 
Here, an additional site is introduced under which mirror fermions are charged that couple with BSM fermions and makes them massive (i.e. they are vector-like).
The gauge structure is identical to the previous model, with the low energy effective theory again being a 2HDM. 
The vector-like fermions now include doublets and singlets, replicating the structure of the SM, 
i.e. the new vector-like fermions are $Q^\prime, L^\prime, {U^c}^\prime, {D^c}^\prime, {E^c}^\prime, {\nu^c}^\prime$. 

\item[]
{The littlest-Higgs with T-parity by Low (Ref.~\cite{Low:2004xc})}: 
Of the two choices of $\cal{G}$ presented in this paper, we consider the group $SU(5)_l \times SO(5)_r/SO(5)_v$.
The low-energy effective theory is a 2HDM plus a singlet complex scalar. 
The mass of the extra doublet is controlled by the $\epsilon_1$ parameter, the coefficient of the plaquette operator. 
For $\epsilon_1 \ll 1$, both Higgs doublets are light, while for $\epsilon_1 \sim 4\pi$, the extra Higgs doublet mass is of the order of $4\pi f$ (i.e. 10~TeV).  
The new fermions are two singlet vector-like quarks and one doublet vector-like quark, all up-type with EM charge $+2/3$.   

\item[]
{A little-Higgs model by Kaplan and Schmaltz (Ref.~\cite{Kaplan:2003uc})}:
The global symmetry structure is $[SU(4)/SU(3)]^4$ with $SU(4)\times U(1)$ gauged. 
The low-energy effective theory is a 2HDM. 
The new fermions are two up-type vector-like quark singlets with EM charge $+2/3$. 

\item[]
{Variation of the littlest-Higgs by Low, Skiba and Smith (LSS, Ref.~\cite{Low:2002ws})}:
The global symmetry structure is taken to be $SU(6)/Sp(6)$, in which $[SU(2)]^2$ is gauged whose diagonal sub-group is identified with the SM $SU(2)$ gauge-group.
The $U(1)_Y$ is not contained in the $SU(6)$. The low-energy effective theory is a 2HDM. 
The new fermions are one vector-like quark doublet with $Y=1/6$, and two vector-like quark singlets which are one up-type with EM charge $+2/3$ and one down-type with EM charge $-1/3$.

\end{itemize}
As a concrete example of the phenomenology of new scalars and vector-like fermions in little-Higgs models, 
we focus here on the last model listed above by Low, Skiba and Smith (LSS, Ref.~\cite{Low:2002ws}), whose
effective theory at the TeV scale is a 2HDM with heavy vector-like fermions and heavy vector-bosons. 
Here, we study in detail the scalar sector of the LSS little-Higgs model, 
after requiring that the lightest CP-even neutral state has the properties
of the $125~$GeV state observed at the LHC.
Since the couplings of this state to SM states measured at the LHC are close to the SM values, 
it will place non-trivial constraints on the parameter space of the model. 
Whether the model evades these constraints successfully, and if it does, 
what features the surviving region of parameter-space has, are the main focus of this work.
This can be relevant to inform future searches at the LHC of the LSS model, and also similar little-Higgs 
models with a 2HDM structure.

We list next a few other studies related to our work.   
A comprehensive discussion on the theory and phenomenology of general 2HDMs is in 
Ref.~\cite{Branco:2011iw} and references therein. 
Constraints on 2HDMs after the LHC Higgs discovery are discussed in Refs.~\cite{dumont2HDM.BIB}.
A model which naturally realizes the 2HDM ``alignment limit'' is studied in Ref.~\cite{Dev:2014yca}.
Fine-tuning in the little-Higgs context is discussed in Refs.~\cite{fineTuning.BIB}.   
Some other studies investigating the compatibility of the little-Higgs with the 
properties of the 125~GeV state observed at the LHC are in Refs.~\cite{littleHiggsLHC.BIB}.

The paper is organized as follows:
In Sec.~\ref{eff2HDM.SEC} we study an effective 2HDM Lagrangian of the type generated in the LSS model.   
In Sec.~\ref{LSS.SEC} we give in detail all theoretical details of the LSS model relevant to our work.
We study the phenomenology of this model in detail in Sec.~\ref{LSSpheno.SEC}, including 8~TeV LHC constraints,
precision electroweak constraints, and present the effective couplings of neutral scalars to two gluons, 
and various branching-ratios of the scalars to SM final-states.
We offer our conclusions in Sec.~\ref{Concl.SEC}.
The 1-loop expressions for the neutral scalar (both CP-even and CP-odd) coupling to two gluons or two photons due to SM or BSM fermions are given in App.~\ref{phiggAA.APP}.
We list some sample points that satisfy the constraints we have considered in App.~\ref{SampPts.APP}.

\section{Effective 2HDM Analysis}
\label{eff2HDM.SEC}

Here we analyze a 2HDM with only certain terms nonzero in the potential, namely,
\beq
{\cal V}_{LSS} = m_1^2 |\phi_1|^2 + m_2^2 |\phi_2|^2 + (b^2 \phi_1^T \cdot \phi_2 + {\rm h.c.}) + \lambda_5^\prime |\phi_1^T \cdot \phi_2|^2 \ , 
\label{2HDMeffL.EQ}
\eeq
where $\phi_1$ and $\phi_2$ are $SU(2)$ doublet scalars with hypercharge $+1/2$ and $-1/2$ respectively,
and $\phi_1^T \cdot \phi_2 \equiv \phi_1^T i\sigma^2 \phi_2$ denotes the antisymmetric product of the fields.
This 2HDM structure is generated in the LSS model, as we explain in detail in Sec.~\ref{LSS.SEC}. 
The introduction of fermions into this 2HDM and the gauge structure are also discussed in detail in that section. 

We require that at the minimum of ${\cal V}_{LSS}$, the vacuum expectation values (VEV) of $\phi_1$ and $\phi_2$ are nonzero, breaking the electroweak symmetry spontaneously down to $U(1)_{EM}$. 
As noted in Ref.~\cite{Low:2002ws}, a sufficient condition for this is $m_{1,2}^2 > 0$ (to prevent VEVs running away to $\infty$) and $(m_1^2 m_2^2 - b^4) < 0$.
The input Lagrangian parameters must be such that these conditions hold, in which case, at the minimum we can take
$\left< \phi_1 \right> = (1/\sqrt{2})(0\ v_1)^T$ and $\left< \phi_2 \right> = (1/\sqrt{2})(v_2\ 0)^T$.
We have\footnote{In Ref.~\cite{Gopalakrishna:2015wwa} we take a different convention of $\phi_1$ and $\phi_2$ both having a hypercharge of $1/2$,  
  and a definition of $\tan \beta = v_2/v_1$.}
\beq
\tan\beta \equiv v_1/v_2 = \sqrt{m_2^2/m_1^2} \ ,
\label{tbeta.EQ}
\eeq
and $v \equiv \sqrt{v_1^2 + v_2^2} = 246~$GeV is fixed from observables.
We find that $v$ is determined as 
\beq
v^2 = \frac{2}{\lambda_5^\prime} \frac{(1+\tan^2\beta)}{\tan\beta} \left(b^2 - m_1^2 \tan\beta\right) \ .
\label{fasv.EQ}
\eeq
 
Taking the fields around the true minimum, we separate the fields in $\phi_{1,2}$, expanding the SU(2) structure, as
\beq
\phi_1 =  \bmat \phi_1^+ \\ (v_1 + \rho_1 + i\eta_1)/\sqrt{2} \emat  \ ; \qquad \phi_2 =  \bmat (v_2 + \rho_2 + i\eta_2)/\sqrt{2} \\ \phi_2^- \emat  \ ,
\eeq
where, $\rho_{1,2}$ are CP-even fields, while $\eta_{1,2}$ are CP-odd. One combination of $\eta_{1,2}$, labelled $G$, is massless and is absorbed in unitary gauge to become the longitudinal $Z_\mu$,
and one combination of $\phi_{1,2}^+$ ($\phi_{1,2}^-$), labelled $G^+$ ($G^-$), is massless and is absorbed into $W^+_\mu$ ($W^-_\mu$).
This leaves two (real) CP-even scalars ($h$, $H$), one (real) CP-odd scalar ($A$) and one (complex) charged scalar ($H^\pm$) as physical states.
We have
\beq
\bmat \rho_1 \\ \rho_2 \emat = \bmat \cos{\alpha} & \sin{\alpha} \\ -\sin{\alpha} & \cos{\alpha} \emat \bmat h \\ H \emat \ ; \qquad
\bmat \eta_1 \\ \eta_2 \emat = - \bmat \cos{\beta} & -\sin{\beta} \\ \sin{\beta} & \cos{\beta} \emat \bmat A \\ G \emat \ , 
\eeq
with
\beq
\tan{(2 \alpha)} = \frac{- 2 (b^2 - v_1 v_2 \lambda_5^\prime)}{(m_2^2 + \lambda_5^\prime v_1^2/2) - (m_1^2 + \lambda_5^\prime v_2^2/2)} \ ; \quad
\tan{(2 \beta)} = \frac{-2 b^2}{(m_2^2 + \lambda_5^\prime v_1^2/2) - (m_1^2 + \lambda_5^\prime v_2^2/2)} \ .
\label{t2albe.EQ}
\eeq
If $\alpha$ and $\beta$ are solutions to Eq.~(\ref{t2albe.EQ}), so are ($\alpha+\pi/2$) and ($\beta+\pi/2$); among these choices, we pick that which ensures $m_h < m_H$ and $m_G = 0$.  
We define $c_\alpha \equiv \cos{\alpha}$, $s_\alpha \equiv \sin{\alpha}$, and, $c_\beta \equiv \cos{\beta}$, $s_\beta \equiv \sin{\beta}$.
The rotation that takes $(\phi_1^\pm , \phi_2^\pm)$ to $(H^\pm , G^\pm)$ is identical to the CP-odd scalars above, i.e. rotation by angle $\beta$.
The mass eigenvalues are given by
\bea
\label{mass-terms}
m_A^2 = 2 b^2/\sin{(2\beta)} \ ; \qquad m_{H^\pm}^2 = m_A^2 - \lambda_5^\prime v^2/2 \ , \nonumber \\ 
m^2_{H,h} = \frac{1}{2} \left[ m_A^2 \pm \sqrt{m_A^4 - 4 (m_A^2 -m_{H^\pm}^2) m_{H^\pm}^2 \sin^2{(2\beta)} }  \right] \ ,
\eea
in agreement with Ref.~\cite{Low:2002ws}.
We identify the lighter CP-even scalar state ($h$) to be the 125~GeV state that has been observed at the LHC.  

The CP-even scalar couplings to $W^+W^-$ is given by
\beq
{\cal L}_{hW^+W^-} = \frac{g^2 v}{2} W_\mu^+ W^{-\, \mu} \left[ \sin{(\beta - \alpha)} h + \cos{(\beta-\alpha)} H  \right] \ .
\label{CPeWW.EQ}
\eeq
Similarly, the $hZZ$ coupling is proportional to $\sin{(\beta-\alpha)}$, 
and $HZZ$ coupling to $\cos{(\beta-\alpha)}$. 
The $hhW^+W^-$ is exactly SM like.
We denote $c_{\theta} \equiv \cos{\theta}$ and $s_{\theta} \equiv \sin{\theta}$.
The $h A Z$ coupling is given by
\beq
{\cal L}^{AZh} =  \frac{g}{2 \cos \theta_W} c_{\beta - \alpha} \left( Z^\mu A \partial_\mu h -  Z^\mu h \partial_\mu A \right) \ .
\eeq
The $W^\pm H^\pm h $ couplings are given by
\beq
{\cal L}^{W^{\pm} H^{\pm} h} =  - i \frac{g}{2} c_{\beta - \alpha} \left( W^{+ \mu} h \partial_\mu H^-  - W^{+ \mu} H^- \partial_\mu h  + h.c.\right) \ .
\eeq
The $Hhh$ coupling can be obtained from Eq.~(\ref{2HDMeffL.EQ}) as
\beq
{\cal L}^{H h h} = -\frac{ \lambda_5^\prime v}{2} \left( 2 c_{2 \alpha} c_{ \beta + \alpha} - s_{2 \alpha} s_{ \beta + \alpha} \right) h h  H \ .
\eeq
Finally the $HH^+ H^-$ coupling can be obtained from Eq.~(\ref{2HDMeffL.EQ}) as
\beq
{\cal L}^{H H^+ H^-} = -\frac{ \lambda_5^\prime v}{2} s_{2 \beta} s_{\beta + \alpha} H H^+ H^- \ .
\eeq
The $H^\pm W_\mu^\mp A^\mu$, $H^\pm W_\mu^\mp Z^\mu$, $hA_\mu Z^\mu$, $HA_\mu Z^\mu$, $H Z_\mu h$ are all zero.

\begin{figure}
\centering
\includegraphics[width=0.32\textwidth]{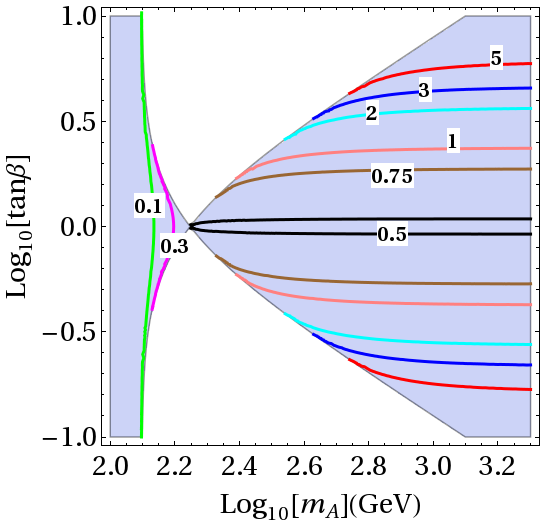}
\includegraphics[width=0.32\textwidth]{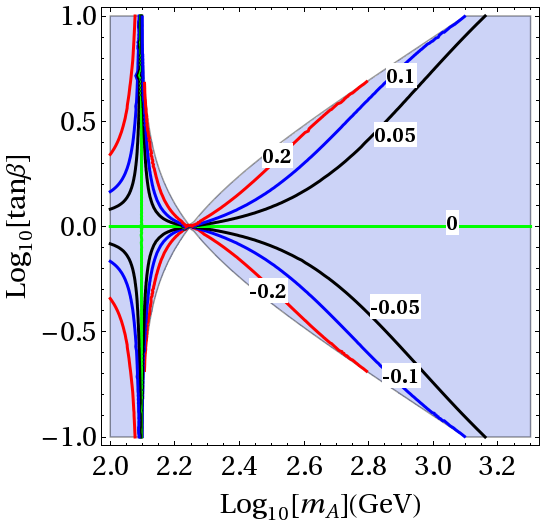}
\includegraphics[width=0.32\textwidth]{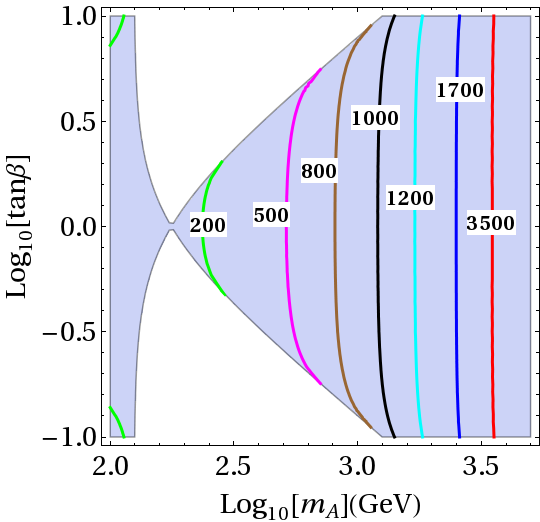}
\caption{Contours of $\lambda_5'$  (left), $\cos(\beta- \alpha)$ (middle) and $m_{H^{\pm}}$ (right) in GeV with $m_{h}=125$~GeV.
  The part of the parameter space for which $\lambda_5'$ is real is shown in the shaded (light-blue) region.
  }
\label{lam-cba-ma-tanb-lss}
\end{figure}

The effective 2-Higgs doublet model with the structure of Eq.~(\ref{2HDMeffL.EQ})
is given in terms of the four parameters $m_1^2$, $m_2^2$, $b^2$ and $\lambda_5^\prime$. 
The constraints on these are $m_h = 125~$GeV and $v\approx 246$~GeV. 
This leaves two parameters free, which we trade for $m_{A}$ and $\tan \beta$. 
In Fig.~\ref{lam-cba-ma-tanb-lss}, we show contours of $\lambda_5^\prime$, $\cos(\beta-\alpha)$ and $m_{H^\pm}$ in the $m_{A}$-$\tan \beta$ plane.
The unshaded (white) region in Fig.~\ref{lam-cba-ma-tanb-lss} is unphysical as $\lambda_5'$ develops an imaginary part.
The $hWW$ and $hZZ$ couplings are constrained by the LHC data to be SM-like to a few tens of percent, which implies $s_{\beta-\alpha} \approx 1$ from Eq.~(\ref{CPeWW.EQ}).  
We discuss this in full detail in Sec.~\ref{LSSpheno.SEC} in the context of the LSS model.
We discuss next the LSS model details, and give expressions for the 2HDM effective parameters in terms of the 
LSS model parameters. 

\section{The Low-Skiba-Smith (LSS) model}
\label{LSS.SEC}

The LSS model is introduced and discussed in detail in Ref.~\cite{Low:2002ws}, 
and we do not repeat all the details here, but rather concentrate on aspects important for our focus here.  
We require the BSM vector bosons to be somewhat heavier in order to avoid precision electroweak constraints.
The effective theory at the TeV scale is then a 2HDM with vector-like fermions and somewhat heavier vector bosons. 
Our main focus will be on the phenomenology of the scalars including the effect of vector-like fermions on them. 
In this section we give all relevant details necessary to our work. 
In some places our notation differs from that in Ref.~\cite{Low:2002ws}.

The global symmetry structure in the LSS model is $SU(6)/Sp(6)$.
The pNGBs $\pi^a$ are contained in the $\Sigma$ as
$$ \Sigma = \exp \{  i \pi^a X^a/f \} \left< \Sigma \right> \ , $$
where the $SU(6)$ is broken down to $Sp(6)$ by an antisymmetric condensate
$$ \left< \Sigma \right> = \bmat 0 & -\mathbbm{1}_{3\times 3} \\ \mathbbm{1}_{3\times 3} & 0 \emat \ , $$
and the $X^a$ are the broken generators.
$f$ sets the scale of the theory, and we take it as an input to the effective theory.
($f$ could be dynamically generated by the UV completion which we will not specify here.)
The 2HDM fields are contained in the pNGB $\pi^a$ that remain light after turning on the gauge and Yukawa couplings.
We have~\cite{Low:2002ws}
\beq
\pi^a X^a \supset \bmat
\begin{matrix} 0 & 0 \\ 0 &  0 \end{matrix} & \phi_2 & \begin{matrix} 0 & s \\ -s &  0 \end{matrix} & \phi_1 \\
\phi_2^\dagger & 0 & -\phi_1^T & 0 \\
\begin{matrix} 0 & -s^* \\ s^* &  0 \end{matrix} & -\phi_1^* & \begin{matrix} 0 & 0 \\ 0 &  0 \end{matrix} & \phi_2^* \\
\phi_1^\dagger & 0 & \phi_2^T & 0 \emat \ ,
\eeq
where we show the (light) pNGB two Higgs-doublets $\phi_1$ and $\phi_2$, and also the (heavy) singlet $s$.
Integrating-out the heavy $s$ generates quartic couplings of the $\phi_1$ and $\phi_2$.

Collective symmetry breaking is ensured in the gauge sector by gauging $SU(2)_1\otimes SU(2)_2$ with gauge couplings $g_1$ and $g_2$.
The $SU(2)$ generators are taken as $Q_1^a$ ($Q_2^a$) with the Pauli matrices $\sigma^a$ in the uppermost (lowermost) $2\times 2$ block and zeros elsewhere;
its diagonal subgroup is identified with the SM $SU(2)$ gauge-group with gauge coupling $g$.
Hypercharge ($U(1)_Y$) is not contained in the $SU(6)$. 
Denoting the hypercharge transformation as $\Sigma \to e^{(i \epsilon Y_L)} \Sigma e^{-(i \epsilon Y_R)} $,
we can take $Y_L^\Sigma = {\rm diag}(0_{2\times 2},\ 1,\ 0_{2\times 2},\ 0) $ and $Y_R^\Sigma = {\rm diag}(0_{2\times 2},\ 0,\ 0_{2\times 2},\ 1) $.  
This results in the hypercharge assignments $Y_{\phi_1} = +1/2$ and $Y_{\phi_2} = -1/2$. 
$\left<\Sigma\right>$ breaks $U(1)_1\otimes U(1)_2$ with gauge couplings $g'_1$ and $g'_2$ down to the diagonal,
which is identified with the SM $U(1)_Y$. 
The light SU(2) and U(1) gauge bosons (massless before EWSB) are identified with the SM $W_\mu$ and $B_\mu$ respectively,
and we denote the corresponding heavy gauge bosons as $W^\prime_\mu$ and $B^\prime_\mu$. 
We have 
\beq
1/g^2 = 1/g_1^2 + 1/g_2^2 \ , \quad 1/g^{\prime\, 2} = 1/g_1^{\prime\, 2} + 1/g_2^{\prime\, 2} \ .  
\label{ggp12.EQ}
\eeq

In the fermion sector also, collective symmetry breaking is ensured by a special structure of the Yukawa couplings~\cite{Low:2002ws}
\beq
    {\cal L}_{Yuk} = y_1 f \bmat Q' & \psi_1 & (i\sigma^2 Q)^T & 0 \emat (\Sigma)^* \bmat 0 \\ t^c \emat
    + y_2 f \bmat 0 & 0 & Q^T & 0 \emat (\Sigma) \bmat i\sigma^2 Q^{\prime\, c} \\ \psi_1^c \\ 0 \\ \psi_2^c \emat + {\rm h.c.} \ .
\label{tYuk.EQ}
\eeq
In this model, in addition to the 3$^{rd}$ generation SM Weyl-fermions $Q, t^c$, new vector-like Weyl-fermion pairs are introduced. 
The new fermions are one vector-like quark doublet Weyl-fermion pair $Q',{Q'}^c$ with $Y=1/6$ and EM charge $2/3$,
one vector-like up-type quark singlet $\psi_1,\psi_1^c$ with EM charge $\pm 2/3$, and one vector-like down-type quark singlet $\psi_2,\psi_2^c$ with EM charge $\mp 1/3$.
We expand the SU(2) structure as $Q=(t,b)^T$, $Q'=(t',b')^T$ and ${Q'}^c=(-{b'}^c,{t'}^c)^T$.
Expanding the Yukawa couplings and including vector-like fermion masses, we get 
\footnote{
Our notation differs from that in Ref.~\cite{Low:2002ws}, and the translations in going from $LSS\to ours$ is:
$\lambda_i \to y_i$, $v_1 \leftrightarrow v_2$.
We also take this opportunity to correct a few minor typos in Ref.~\cite{Low:2002ws}. 
}
\bea
    {\cal L}^{\rm ferm} \supset && -  y_1  \left(f \psi_1 t^c - i {Q'}^T \phi_2^* t^c - i Q^T \cdot \phi_1 t^c \right)
    + y_2  \left(f Q^T \cdot {Q'}^c + i Q^T \phi_1^* \psi_2^c + i Q^T \phi_2^* \psi_1^c \right) \nonumber \\ 
    && + y_3 f {Q'}^T \cdot {Q'}^c + y_4 f \psi_1^c \psi_1 + y_5 f \psi_2^c \psi_2 + h.c. \ ,
    \label{Lferm.EQ}
\eea
where again, the ``$\cdot$'' represents the anti-symmetric combination of the SU(2) indices.

As shown in Ref.~\cite{Low:2002ws}, the gauge and Yukawa coupling structure above breaks the global SU(6) explicitly, allowing the Higgs to acquire a mass at the loop-level.
But most importantly for the little-Higgs, the breaking is {\em collective}, in that if either gauge coupling or if either Yukawa coupling is set to zero, there is an exactly preserved global symmetry
keeping the Higgs massless.
Therefore, the loop generated Higgs mass must be proportional to a product of both the gauge couplings, and similarly both the Yukawa couplings, which implies that the Higgs mass is finite at 1-loop. 

For generating the bottom mass, we introduce an SU(2) singlet field $b^c$ and write a Yukawa coupling as
\beq
    {\cal L}^b_{Yuk} = 
    -i y^{(1)}_b f \bmat 0 & 0 & Q^T & 0 \emat (\Sigma) \bmat 0 \\ 0 \\ 0 \\ b^c \emat
    + i y^{(2)}_b f \bmat 0 & 0 & (i \sigma_2 Q)^T & 0 \emat (\Sigma)^* \bmat 0 \\ b^c \\ 0 \\ 0 \emat 
    + h.c. \ .
\label{bYuk}
\eeq
Expanding ${\cal L}^b_{Yuk}$  gives
\beq
    {\cal L}^b_{Yuk} \supset y^{(1)}_b  Q^T \phi_1^* b^c - y^{(2)}_b  Q^T \cdot \phi_2 b^c  + {\rm h.c.} \ .
    \label{bYukcomp}
\eeq
The $s$-quark mass is generated in an identical fashion, with the replacement $y_b \to y_s$.
To generate the $c$-quark mass we introduce an SU(2) doublet field $Q_2 = (c,s)^T$ and an SU(2) singlet field $c^c$ and add the following Yukawa term
\beq
    {\cal L}^c_{Yuk} =-i y^{(1)}_c f \bmat 0 & 0 & (i\sigma^2 Q_2)^T & 0 \emat (\Sigma)^* \bmat 0 \\ c^c \emat
    -i y^{(2)}_c f \bmat 0 & 0 & Q_2^T & 0 \emat (\Sigma) \bmat 0 \\ c^c \\ 0 \\ 0 \emat + {\rm h.c.} \ .
\label{cYuk.EQ}
\eeq
Expanding ${\cal L}^c_{Yuk}$ gives
\beq
    {\cal L}^c_{Yuk} \supset y^{(1)}_c  Q_2^T \cdot \phi_1 c^c + y^{(2)}_c  Q_2^T  \phi_2^* c^c  + {\rm h.c.} \ .
    \label{cYukcomp}
\eeq
Similarly, for the $\tau$-lepton we have
\beq
     {\cal L}^\tau_{Yuk} \supset y^{(1)}_\tau  L^T \phi_1^* \tau^c - y^{(2)}_\tau L^T \cdot \phi_2 \tau^c + {\rm h.c.},
    \label{tauYukcomp}
\eeq
where $L$ is the $SU(2)$ doublet lepton with $Y=-1/2$ and $\tau^c$ is the $SU(2)$ singlet lepton with $Y=1$.
We generate masses for the other light SM fermions in an analogous way.
This structure of ${\cal L}^{b,c,s,\tau}_{Yuk}$ and the other light fermions does not implement the little-Higgs mechanism. We do not worry about this in the bottom sector and the other light fermions
as we find that the phenomenologically acceptable parameter-space has $\tan\beta \sim {\cal O}(1)$ for which the Yukawa couplings for these fermions are all small enough that
the fine-tuning they necessitate are not significant.

The Higgs potential generated at 1-loop in the LSS model is that of Eq.~(\ref{2HDMeffL.EQ})~\cite{Low:2002ws}.
We turn next to analyzing the model in detail in terms of the input Lagrangian parameters. 
In particular, the $m_1^2$, $m_2^2$, $b^2$ and $\lambda_5^\prime$ are functions of the input Lagrangian parameters, 
as given below~\cite{Low:2002ws}:
\bea
\lambda_5^\prime &=& \frac{c g_1^2 \left[g_2^2 + (c'/c) y_2^2 \right])}{g_1^2 + g_2^2 + (c'/c)y_2^2} \ , \quad
b^2 = \frac{3 f^2}{8 \pi^2} y_1^2 y_2 (y_3 - y_4) \log{\frac{\Lambda^2}{M_f^2}} \ , \nonumber \\
m^2_{1\, f} &=& \frac{3 f^2}{8 \pi^2} (y_1^2 - y_2^2) (y_3^2 - y_4^2) \log{\frac{\Lambda^2}{M_f^2}} \ , \quad
m^2_{2\, f} = \frac{3 f^2}{8 \pi^2} (y_1^2 y_2^2 + y_2^2 y_5^2 - y_2^2 y_3^2 - y_1^2 y_4^2) \log{\frac{\Lambda^2}{M_f^2}} \ , \nonumber \\
m^2_{1g} &=& m^2_{2g} = \frac{3}{64\pi^2} \left[ 3 g^2 M_g^2 \log{\frac{\Lambda^2}{M_g^2}} + g^{\prime\, 2} M_{g'}^2 \log{\frac{\Lambda^2}{M_{g'}^2}} \right] \ , \quad 
m^2_{1s} = m^2_{2s} = \frac{\lambda_5^\prime}{16 \pi^2} M_s^2 \log{\frac{\Lambda^2}{M_S^2}} \ , 
\label{2HDMfLSS.EQ}
\eea
where $\Lambda$ is the cut-off which is taken to be $4\pi f$. $M_f$ is the heavy vector-like fermion mass-scale.
The heavy gauge-boson masses are $M_g = f\sqrt{(g^2_1 + g^2_2)/2}$ and $M_{g^\prime} = f\sqrt{(g^{\prime\, 2}_1 + g^{\prime\, 2}_2)/2}$.  
The singlet scalar ($s$) mass is $M_s = f \sqrt{c(g_1^2 + g_2^2)+c' y_2^2}$,
where $c$ and $c'$ are $O(1)$ parameters that depend on the UV completion details as explained in Ref.~\cite{Low:2002ws}. 

From Eq.~(\ref{Lferm.EQ}), we can infer the fermion mass matrix after EWSB. The EM charge $+2/3$ and $-1/3$ fermion mass matrices are
\beq
    {\cal L} \supset \bmat t & \psi_1 & t' \emat \bmat
    i y_1 \frac{v_1}{\sqrt{2}} & iy_2 \frac{v_2}{\sqrt{2}} & y_2 f  \\
    -y_1 f & y_4 f & 0 \\
     iy_1 \frac{v_2}{\sqrt{2}} & 0 &  y_3 f 
     \emat \bmat t^c \\ \psi_1^c \\ {t'}^c \emat
     + \bmat b & \psi_2 & b' \emat
     \bmat y_b^{(i)} \frac{v_i}{\sqrt{2}} & i y_2 \frac{v_1}{\sqrt{2}} & y_2 f \\ 0 & y_5 f & 0 \\ 0 & 0 & y_3 f \emat
     \bmat b^c \\ \psi_2^c \\ {b'}^c  \emat  + {\rm h.c.} \ ,
     \label{MudIntB.EQ}
\eeq
where $v_i = \{ v_1, v_2\}$. 
To work out the couplings of the scalars to up-type fermions in the mass basis, we diagonalize Eq.~(\ref{MudIntB.EQ}).
We implement a two-step diagonalization, where first the $f$-dependent terms are diagonalized, and then the $v_{1,2}$ dependent EWSB pieces.
The rotations that diagonalize the $f$ dependent terms are
\bea
\bmat t \\ \psi_1 \\ t^\prime \emat = \bmat c_{23} & 0 & -s_{23} \\ 0 & 1 & 0 \\ s_{23} & 0 & c_{23} \emat \bmat t_0 \\ \psi_1 \\ t_1 \emat ; \ 
\bmat t^c \\ \psi_1^c \\ {t^\prime}^c \emat = \bmat c_{14} & -s_{14} & 0 \\ s_{14} & c_{14} & 0 \\ 0 & 0 & 1 \emat \bmat t_0^c \\ t_1^c \\ {t^\prime}^c \emat ; \ 
\bmat b \\ \psi_2 \\ b^\prime \emat = \bmat c_{23} & 0 & -s_{23} \\ 0 & 1 & 0 \\ s_{23} & 0 & c_{23} \emat \bmat b_0 \\ \psi_2 \\ b_1 \emat, 
\label{fdiagRot.EQ}
\eea
with $s_{23} \equiv \sin\theta_{23} = y_2/(\sqrt{y_2^2 + y_3^2})$, $c_{23} \equiv \cos\theta_{23} = -y_3/(\sqrt{y_2^2 + y_3^2})$, and $s_{14} \equiv \sin\theta_{14} = y_1/(\sqrt{y_1^2 + y_4^2})$. 
After these rotations the mass matrices become
\beq
    {\cal L}^{\rm mass} \supset \bmat t_0 & \psi_1 & t_1 \emat
    \bmat -i {\cal M}^{t}_{11} & i {\cal M}^{t}_{12} & 0 \\ 0 & {\cal M}^{t}_{22} & 0 \\ -i {\cal M}^{t}_{31} & i {\cal M}^{t}_{32} & -{\cal M}^{t}_{33} \emat
    \bmat t_0^c \\ t_1^c \\ t^{\prime\, c}   \emat 
    +  \bmat b_0 & \psi_2 & b_1 \emat
    \bmat {\cal M}^{b}_{11}  & i {\cal M}^{b}_{12}  & 0 \\ 0 & {\cal M}^{b}_{22} & 0 \\ - {\cal M}^{b}_{31}  & i {\cal M}^{b}_{32} & - {\cal M}^{b}_{33} \emat
    \bmat b^c \\ \psi_2^c \\ b^{\prime\, c}   \emat
    + {\rm h.c.} \ .
    \label{Mfdiag.EQ}
\eeq
where $\sqrt{y}_{14} \equiv \sqrt{y_1^2 + y_4^2}$, $\sqrt{y}_{23} \equiv \sqrt{y_2^2 + y_3^2}$ and the elements of the mass matrix are given by
\bea
{\cal M}^{t}_{11} &=& \frac{y_1 (y_3 y_4 v_1 + y_2 y_3 v_2 - y_2 y_4 v_2)}{\sqrt{y}_{14} \sqrt{y}_{23} \sqrt{2}} \ , \qquad
{\cal M}^{t}_{12} = \frac{(y_1^2 y_3 v_1 - y_2 y_3 y_4 v_2 - y_1^2 y_2 v_2)}{\sqrt{y}_{14} \sqrt{y}_{23} \sqrt{2}} \ , \nonumber \\
{\cal M}^{t}_{31} &=& \frac{y_1 (y_2 y_4 v_1 + y_2^2 v_2 + y_3 y_4 v_2)}{\sqrt{y}_{14} \sqrt{y}_{23} \sqrt{2}}  \ , \qquad
{\cal M}^{t}_{32} = \frac{(y_1^2 y_2 v_1 - y_2^2 y_4 v_2 + y_1^2 y_3 v_2)}{\sqrt{y}_{14} \sqrt{y}_{23} \sqrt{2}} \ , \nonumber \\
{\cal M}^{t}_{22} &=& f \sqrt{y}_{14} \ , \qquad   {\cal M}^{t}_{33} = f \sqrt{y}_{23} \ , \nonumber \\
{\cal M}^{b}_{11} &=& c_{23} y^{(i)}_b \frac{v_i}{\sqrt{2}} \ , \qquad
{\cal M}^{b}_{12} = y_2 \frac{v_1}{\sqrt{2}} \ , \qquad
{\cal M}^{b}_{22} = y_5 f  \ , \nonumber \\
{\cal M}^{b}_{31} &=& y^{(i)}_b \frac{v_i}{\sqrt{2}} s_{23} \ , \qquad 
{\cal M}^{b}_{32} = - y_2 \frac{v_2}{\sqrt{2}} s_{23} \ , \qquad
{\cal M}^{b}_{33} = \sqrt{y}_{23} f 
\label{Mvij.EQ}
\eea
To make the mass matrix entries of Eq.~(\ref{Mfdiag.EQ}) real and positive, we perform the following field redefinitions:
$t_0^c = i \tilde t_0^c$, $t_1^c = -i \tilde t_1^c$, $t^{\prime\, c} = - \tilde t^{\prime\, c}$, $\psi_1 = i \tilde \psi_1$, and,
$b_1 = -\tilde b_1$, $\psi_2^c = -i\tilde \psi_2^c$, and $\psi_2 = i\tilde \psi_2$.
After these field redefinitions the new mass matrix entries are just the ${\cal M}^{t,b}_{ij}$ shown in Eq.~(\ref{Mvij.EQ})
(without the $'i'$s and negative signs in the matrix of Eq.~(\ref{Mfdiag.EQ})).
For brevity of notation, in the following, we drop the tilde on the fields and denote the fields $\tilde{\chi}_i$ simply as $\chi_i$. 
The next step is to diagonalize ${\cal M}^t$.  
We achieve this through a bi-orthogonal transformation given by $R_L$ and $R_R$ as 
$(t_0\ \psi_1\ t_1)^T = R_L^T (\hat{t}_0\ \hat{t}_1\ \hat{t}_2)^T$ and 
$(t_0^c\ t_1^c\ t^{\prime\, c})^T = R_R^T (\hat{t}_0^c\ \hat{t}_1^c\ \hat{t}_2^c)^T$,
such that $R_L {\cal M}^t R_R^T \equiv \hat{\cal M}^t$ is diagonal. 
The $\hat{t}_i, \hat{t}_i^c$ are the mass eigenstate fields. 
We do this diagonalization of the $v_{1,2}$ dependent pieces numerically.
We identify $(\hat{t}_0, \hat{t}_0^c)$ as the observed top-quark.
In the bottom sector, we do not diagonalize the $v_i$ proportional off-diagonal terms as they are numerically insignificant due to the small $y_b$, 
and we identify $(b_0, b^c)$ as the observed bottom quark. 
We denote the mass eigenvalues as $m_t, M_{t_2}, M_{t_3}$ in the top sector, 
and $m_b, M_{b_2}, M_{b_3}$ in the bottom sector. 

We turn next to extracting the top-quark Yukawa coupling $y_{htt}$. 
In the original basis, the $htt$ Yukawa coupling is
\beq
    {\cal L}_{htt} = \frac{i\, h}{\sqrt{2}} \bmat t & \psi_1 & t' \emat
    \bmat y_1 c_\alpha & -y_2 s_\alpha & 0 \\ 0 & 0 & 0 \\ -y_1 s_\alpha & 0 & 0 \emat
    \bmat t^c \\ \psi_1^c \\ t^{\prime\, c} \emat + {\rm h.c.} \ .
\eeq
We rewrite this in the basis where the $f$-terms are diagonal.\footnote{In our convention,
we define the top-quark Yukawa coupling $y_{htt}$ as
$ {\cal L}_{htt} = (h/\sqrt{2}) y_{htt} \hat{t}_0 \hat{t}_0^c+ {\rm h.c.} $.
We define this with a positive sign here since our field redefinitions made the fermion mass terms positive.
}
After the field redefinitions above to render the fermion mass matrix real,
the $h$ Yukawa couplings in the fermion basis after the $f$-terms are diagonalized (but $v$-terms still not diagonalized) are
\beq
{\cal L}^{\rm Yuk}_h \supset  \frac{h}{\sqrt{2}} \left[
  y_{00} t_0 t_0^c  + y_{01} t_0 t_1^c  + y_{10} t_1 t_0^c  + y_{11} t_1 t_1^c \right] + {\rm h.c.} \ ,
\label{hYuk.EQ}
\eeq
with 
$y_{00} \equiv (-y_1 c_\alpha c_{14} c_{23} + y_1 s_\alpha c_{14} s_{23} + y_2 s_\alpha s_{14} c_{23})$, 
$y_{01} \equiv (-y_1 c_\alpha s_{14} c_{23} + y_1 s_\alpha s_{14} s_{23} - y_2 s_\alpha c_{14} c_{23})$, 
$y_{10} \equiv (y_1 c_\alpha c_{14} s_{23} + y_1 s_\alpha c_{14} c_{23} - y_2 s_\alpha s_{14} s_{23})$,  
$y_{11} \equiv (y_1 c_\alpha s_{14} s_{23} + y_1 s_\alpha s_{14} c_{23} + y_2 s_\alpha c_{14} s_{23})$.
Including the bi-orthogonal rotations that take us to the mass-basis, the $y_{htt}$ in the model is given as
$ y_{htt} = \left[ y_{00} (R_L)_{00} (R_R)_{00} + y_{01} (R_L)_{00} (R_R)_{01} + y_{10} (R_L)_{01} (R_R)_{00} + y_{11} (R_L)_{01} (R_R)_{01} \right] \ , $
where $(R)_{ij}$ with $i,j = \{0,1,2\}$, is the $(i+1,j+1)$ entry of the rotation matrix $R$.
We define $\kappa_{htt} \equiv y_{htt}/y_{htt}^{SM}$. 
Similarly, the $H$ couplings $y^H_{ij}$ are got from Eq.~(\ref{hYuk.EQ}) by making the change $c_\alpha \to s_\alpha$ and $s_\alpha \to -c_\alpha$. 

The $A$ couplings to fermions can be obtained from 
\beq
{\cal L}_A^{\rm Yuk} \supset \frac{A}{\sqrt 2} \left[ y_1 \left( \cos\beta \, t - \sin\beta \, t' \right) t^c  
                                                      - y_2 \left( \sin\beta \, t \psi_1^c  + \cos\beta \, b \psi_2^c \right) \right] + {\rm h.c.} \ .
\eeq
After diagonalizing the $f$ terms, in the basis of Eq.~(\ref{fdiagRot.EQ}) and after the field redefinitions shown below Eq.~(\ref{Mvij.EQ}), we have
\beq
{\cal L}^{\rm Yuk}_A \supset  \frac{i\, A}{\sqrt{2}} \left[
y_{00}^A t_0 t_0^c  + y_{01}^A t_0 t_1^c  + y_{10}^A t_1 t_0^c  + y_{11}^A t_1 t_1^c \right] + {\rm h.c.} \ ,
\eeq
with 
$y_{00}^A \equiv (y_1 c_\beta c_{14} c_{23} - y_1 s_\beta c_{14} s_{23} - y_2 s_\beta s_{14} c_{23})$, 
$y_{01}^A \equiv (y_1 c_\beta s_{14} c_{23} - y_1 s_\beta s_{14} s_{23} + y_2 s_\beta c_{14} c_{23})$, 
$y_{10}^A \equiv (-y_1 c_\beta c_{14} s_{23} - y_1 s_\beta c_{14} c_{23} + y_2 s_\beta s_{14} s_{23})$,  
$y_{11}^A \equiv -(y_1 c_\beta s_{14} s_{23} + y_1 s_\beta s_{14} c_{23} + y_2 s_\beta c_{14} s_{23})$.
Diagonalizing the $v$ proportional mass terms via the bi-orthogonal rotations $R_L$ and $R_R$ as explained below Eq.~(\ref{Mvij.EQ}),
we have in the mass basis
\beq
{\cal L}^{\rm Yuk}_A \supset  \frac{i\, A}{\sqrt{2}} \left[
y_{00}^A {R_L}_{j0} {R_R}_{k0} + y_{01}^A {R_L}_{j0} {R_R}_{k1} + y_{10}^A {R_L}_{j1} {R_R}_{k0} + y_{11}^A {R_L}_{j1} {R_R}_{k1} \right] \hat{t}_j \hat{t}_k^c + {\rm h.c.} \ . 
\eeq
For notational brevity, henceforth, we denote $\hat t_0$ simply as $t$, and the heavier EM charge $2/3$ fermions as $t_2$ and $t_3$.

The $b$-quark mass term and coupling to scalars can be derived from Eq.~(\ref{bYukcomp}) as
\beq
    {\cal L} \supset \frac{c_{23}}{\sqrt{2}} \left[v \left(y_b^{(1)} s_\beta + y_b^{(2)} c_\beta \right) + h \left(y_b^{(1)} c_\alpha - y_b^{(2)} s_\alpha \right) + H \left(y_b^{(1)} s_\alpha + y_b^{(2)} c_\alpha \right)
      + i A \left(y_b^{(1)} c_\beta - y_b^{(2)} s_\beta \right) \right] b_0 b^c + {\rm h.c.} \ .
    \label{LhHAbb.EQ}
\eeq
The observed $b$-quark mass is thus identified to be $\hat{m}_b \equiv v (y_b^{(1)} s_\beta + y_b^{(2)} c_\beta )/\sqrt{2}$.
Analogous mass and coupling expressions apply to the $\tau$ with the replacement $y_b \to y_\tau$.
The $c$-quark mass term can be obtained from Eq.~(\ref{cYukcomp}) as
$m_c = ( y^{(1)}_c s_\beta + y^{(2)}_c c_\beta ) v/\sqrt{2}$.
The $s$-quark mass is obtained by the replacement $y_c \to y_s$.

The $H^\pm t b$ couplings can be obtained as
\beq
    {\cal L}^{Yuk} \supset H^+ \left( y^+_{00} b_0 t_0^c + y^+_{01} b_0 t_1^c + y^+_{10} b_1 t_0^c + y^+_{11} b_1 t_1^c \right) +
        H^- \left(y^-_{00} t_0 b^c + y^-_{10} t_1 b^c + y^-_{02} t_0 \psi_2^c + y^-_{12} t_1 \psi_2^c  \right) + {\rm h.c.} \ ,
 \label{lHctb.EQ}
\eeq
where
$y^+_{00} = (y_1 s_\beta s_{23} c_{14} - y_1 c_\beta c_{23} c_{14} + y_2 s_\beta c_{23} s_{14})$,
$y^+_{01} = (y_1 s_\beta s_{23} s_{14} - y_1 c_\beta c_{23} s_{14} - y_2 s_\beta c_{23} c_{14})$,
$y^+_{10} = (-y_1 s_\beta c_{23} c_{14} - y_1 c_\beta s_{23} c_{14} + y_2 s_\beta s_{23} s_{14})$,
$y^+_{11} = (-y_1 s_\beta c_{23} s_{14} - y_1 c_\beta s_{23} s_{14} - y_2 s_\beta s_{23} c_{14})$,
$y^-_{00} = [(-y_b^{(1)} c_\beta + y_b^{(2)} s_\beta) c_{23}]$,
$y^-_{10} = [(y_b^{(1)} c_\beta - y_b^{(2)} s_\beta) s_{23}]$,
$y^-_{02} = (-y_2 c_\beta c_{23})$,
$y^-_{12} = (y_2 c_\beta s_{23})$.
The rotations $R_L, R_R$ that diagonalize the $v_{1,2}$ proportional off-diagonal terms are then applied on these, which we do not explicitly show. 
The $H^\pm cs$ and $H^\pm \tau\nu$ couplings can be obtained as
\beq
{\cal L}^{Yuk} \supset \left(y^{(1)}_c c_\beta - y^{(2)}_c s_\beta \right) H^+ s c^c 
+ \left(- y^{(1)}_s c_\beta + y^{(2)}_s s_\beta \right) H^- c s^c 
+ \left(- y^{(1)}_\tau c_\beta + y^{(2)}_\tau s_\beta \right) H^- \nu \tau^c + {\rm h.c.} \ .
\eeq

\section{Phenomenological Analysis of the LSS Model}
\label{LSSpheno.SEC}

To recapitulate, the LSS model has 12 Lagrangian parameters, $f, g_1, g_2, g'_1, g'_2, y_1, y_2, y_3, y_4, y_5, c, c'$.
We impose the constraints in Eq.~(\ref{ggp12.EQ}) and take $g_1$ and $g'_1$ as independent and $g_2$ and $g'_2$ as determined by these.
This reduces the number of independent parameters to 10. 
Furthermore, we fix $v=246~$GeV and determine $f$ in terms of this and other parameters using 
Eqs.~(\ref{fasv.EQ})~and~(\ref{2HDMfLSS.EQ}). 
At this stage we are then left with 9 input parameters, namely, $g_1$, $g'_1$, $y_1$, $y_2$, $y_3$, $y_4$, $y_5$, $c$, $c'$.
Since the relations between the input parameters and the observables are complicated and not easily invertible analytically,
we scan over these 9 parameters and ask which regions, if any, satisfy experimental constraints.
We give more details on the scan and the results below. 

We list the relevant experimental constraints in Table~\ref{constrVals.TAB}, which are roughly in the 2~to~3$~\sigma$ range.  
The references for the measurements are also shown.
\begin{table}
\protect\caption{The experimental constraints at about the $2~{\rm to}~3~\sigma$ level.
  \label{constrVals.TAB}}
\begin{centering}
\begin{tabular}{|c|c|c|}
\hline 
Quantity & Constraint & Reference\tabularnewline
\hline 
\hline 
Top mass (MSbar) & $158<m_{t}^{\overline{MS}}<168.7$~GeV & Ref.~\cite{Alekhin:2012py} \tabularnewline
\hline 
Higgs VEV & $v\equiv 246$~GeV & \tabularnewline
\hline 
Higgs mass & $123<m_{h}<127$~GeV & Ref.~\cite{Aad:2015zhl} \tabularnewline
\hline 
Higgs Yukawa & $0.63<|\kappa_{htt}|<1.2$ & Table 15 of Ref.~\cite{hCouplComb.BIB} \tabularnewline
\hline 
$hW^{+}W^{-}$ coupling & $|\cos(\beta-\alpha)|<0.4$ & Table 15 of Ref.~\cite{hCouplComb.BIB} \tabularnewline
\hline 
VLQ mass & $M_{t',\, b'}>750$~GeV & Refs.~\cite{TheATLAScollaboration:2013oha},~\cite{TheATLAScollaboration:2013sha} \tabularnewline
\hline 
\end{tabular}
\par\end{centering}
\end{table}
We match the top mass in the LSS model to the $\overline{MS}$ mass shown in the table.
Higgs couplings measured at the LHC so far largely agree with the SM, at least to about a few tens of percent.
As already mentioned, since we identify the scalar state observed at the LHC to be the lighter CP-even state $h$,
the magnitude of the $hVV$ coupling (with $V=\{W_\mu^\pm, Z_\mu \}$) is constrained to be close to the SM coupling at the few tens of percent level.
From Eq.~(\ref{CPeWW.EQ}), we see that to satisfy this constraint, it is sufficient that $|\sin(\beta-\alpha)| \approx 1$. 
This will be realized in the so called ``decoupling limit''~\cite{Gunion:2002zf}, or more generally in the ``alignment limit''~\cite{Bhattacharyya:2015nca}.
For the alignment limit to hold, we need $(\alpha - \beta) \approx \pm \pi/2$.
Although it is common to allow only the positive sign of the $hVV$ coupling (i.e. same sign as in the SM), the data so far does not fix the sign.
The $h\to WW,ZZ$ decays are dominated by the tree-level amplitude and is thus largely insensitive to the sign of the $hVV$ coupling.
However, $h\to\gamma\gamma$ occurs at loop-level and is indeed sensitive to the sign due to interference between the gauge-boson and top-loops,
but only to the relative sign between the $hVV$ and $htt$ couplings.  
Thus, both possibilities remain, i.e. $hVV$ and $htt$ both positive, or alternately both negative.
Therefore we allow both these possibilities in realizing the alignment limit; when $hVV$ is negative, we also demand that $htt$ be negative.  
The $hVV$ coupling constraint we pick and show in Table~\ref{constrVals.TAB} is for the case when no new contributions enter into the $hgg$ loop;
although not strictly true in the model due to the presence of vector-like fermions, this will be a good approximation for heavy vector-like quark masses, and 
our main reason for picking this is that it will lead to a more conservative bound.
The sizes of deviations of the $h$ couplings to SM states due to vector-like fermion contributions at 1-loop are discussed in Ref.~\cite{Ellis:2014dza}.  
We impose the direct LHC limit $M_{t',b'} > 750~$GeV on the vector-like quarks (VLQ) as shown. 
 
\begin{figure}
\centering
\includegraphics[width=0.32\textwidth]{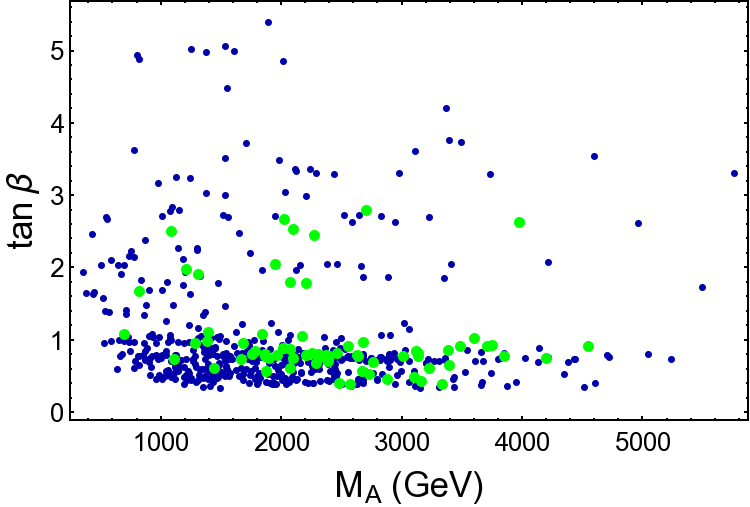}
\includegraphics[width=0.32\textwidth]{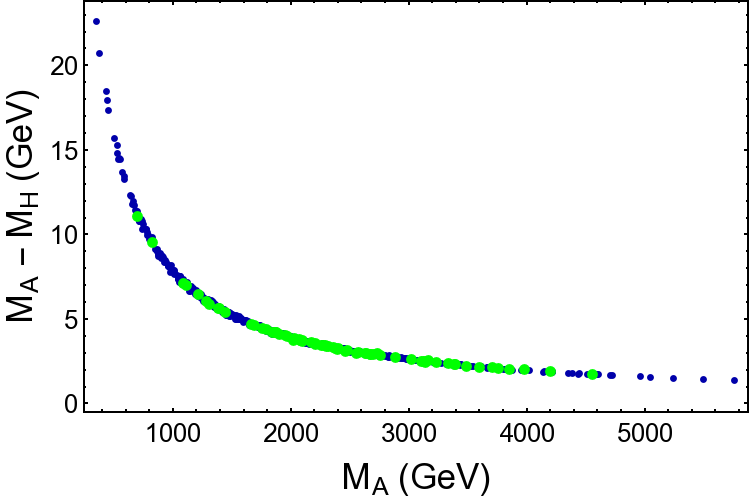}
\includegraphics[width=0.32\textwidth]{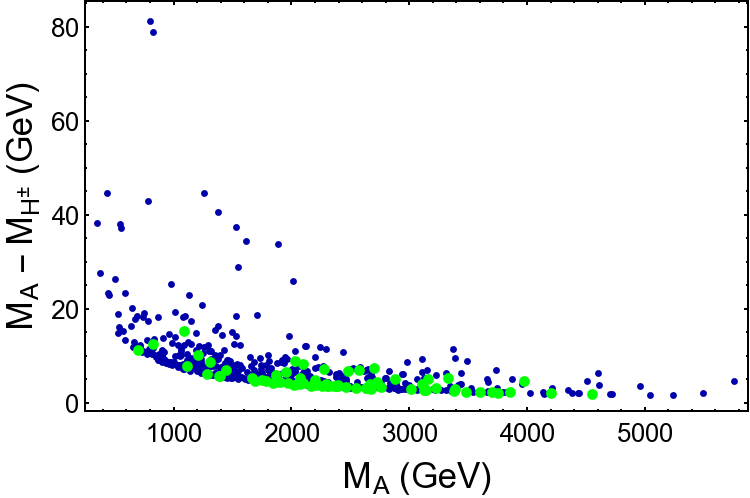}
\caption{The scalar masses and $\tan\beta$ for the LSS model
  for the points that satisfy the direct experimental constraints discussed in the text (blue dots), and in addition the precision electroweak constraints (green dots).}
\label{dots2HDM.FIG}
\end{figure}

\begin{figure}
\centering
\includegraphics[width=0.32\textwidth]{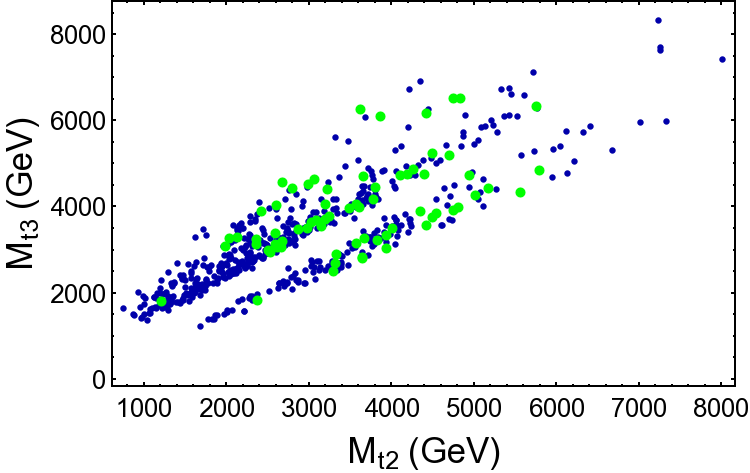} 
\includegraphics[width=0.32\textwidth]{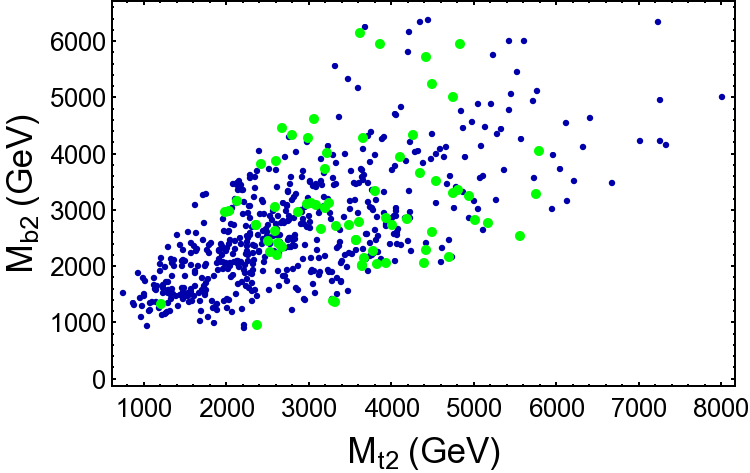} 
\includegraphics[width=0.32\textwidth]{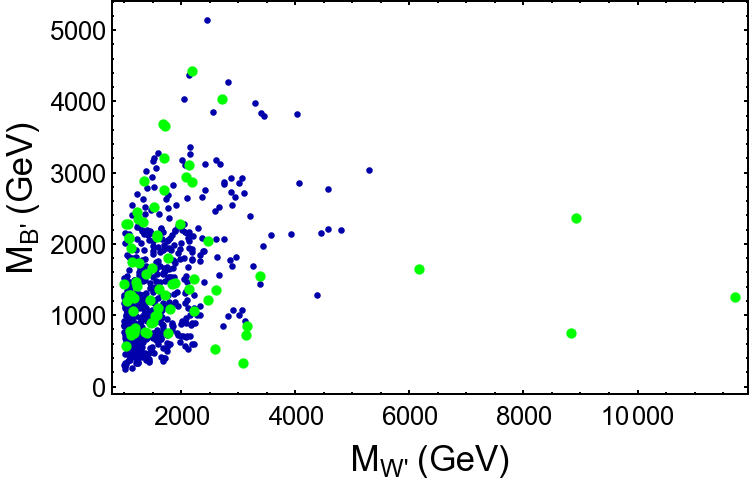}  
\caption{The heavy vector-like fermion and vector-boson masses. The color-coding of the dots is as in Fig.~\ref{dots2HDM.FIG}. 
}
\label{dotsMfMG.FIG}
\end{figure}

\begin{figure}
\centering
\includegraphics[width=0.45\textwidth]{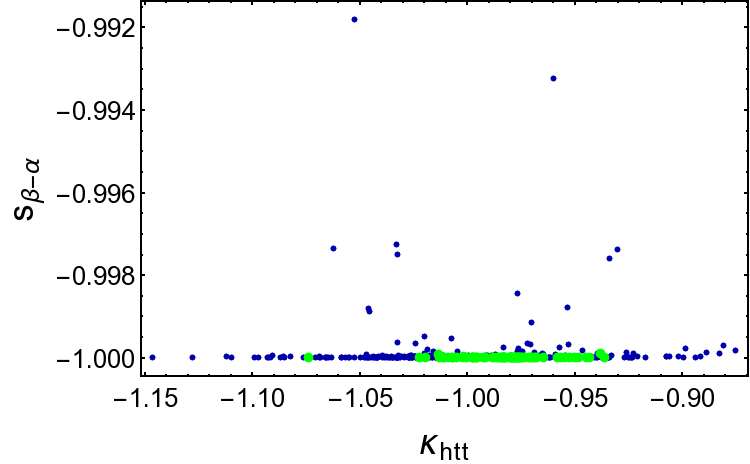}\hspace{1em}%
\caption{$\kappa_{htt}$ and $s_{\beta-\alpha}$, which are the ratios of the $htt$ and $hVV$ couplings to the corresponding SM values, with $VV=\{W^+W^-,ZZ\}$. The color-coding of the dots is as in Fig.~\ref{dots2HDM.FIG}.}
\label{dotskhttsbemal.FIG}
\end{figure}

\begin{figure}
\centering
\includegraphics[width=0.45\textwidth]{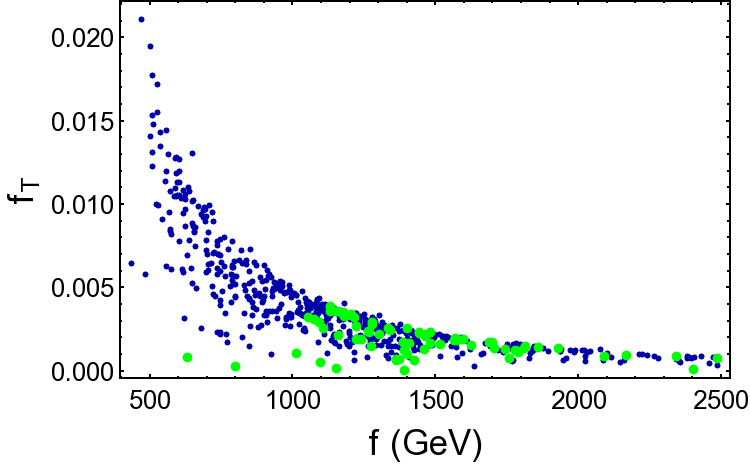}\hspace{1em}%
\caption{The fine-tuning $f_T$ as a function of $f$ in the LSS model. The color-coding of the dots is as in Fig.~\ref{dots2HDM.FIG}.}
\label{dotsffT.FIG}
\end{figure}

Our goal is to scan the 9-dimensional parameter space detailed above in order to find regions where the experimental constraints are all satisfied, and in these regions study the
scalar, fermion and vector-boson sectors of the LSS model.
To systematically sample the 9-dimensional parameters space with the required granularity and identify the experimentally allowed points is computationally too demanding. 
Therefore we randomly sample the parameter space, and for each sample implement a {\em steepest descent} algorithm to minimize the $\chi^2$ cost-function given by  
\beq
\chi^2 \equiv \frac{(m_h - \hat m_h)^2}{\sigma_{m_h}^2} + \frac{(m_t - \hat m_t)^2}{\sigma_{m_t}^2} + \frac{(|\kappa_{htt}|-\hat\kappa_{htt})^2}{\sigma_{htt}^2} + \frac{(c_{\beta-\alpha} - \hat c_{\beta-\alpha})^2}{\sigma_{c_{\beta-\alpha}}^2} \ , 
\eeq
where we choose $\hat m_h = 125~$GeV, $\hat m_t = 163.3~$GeV, $\hat\kappa_{htt} = 1$, $\hat c_{\beta-\alpha} = 0$, 
and the corresponding standard-deviations to be $\sigma_{m_h} = 3$~GeV, $\sigma_{m_t} = 5.4~$GeV, $\sigma_{htt} = 0.25$, $\sigma_{c_{\beta-\alpha}} = 0.2$.  
We start the scan with a random point in the 9-dimensional space and compute the $\chi^2$ and its partial derivative with respect to each of the 9 parameters.
Using the partial derivatives, the normal vector to the $\chi^2$ function at this point is computed and an infinitesimal step in a direction opposite to the normal is taken to get the updated point
with a lower $\chi^2$.
This point is then made the new starting point and the above process iterated till the (local) minimum of the $\chi^2$ is reached.
If this local minimum happens to have a $\chi^2 < 10$, we keep this as a good point; if not, we discard this point and start with another random point.
In this manner, we accumulate a list of points in the 9-dimensional space with $\chi^2 < 10$.
We further cut down this sample to only keep points which satisfy the following additional criteria:
the experimental constraints of Table~\ref{constrVals.TAB} (with $|\kappa_{htt}|$ in the range shown), $\kappa_{htt}$ and $s_{\beta-\alpha}$ are the same sign,
all vector-like quarks ($t'$ and $b'$) masses above $750$~GeV, and $M_{W^\prime} > 1000~$GeV.\footnote{We do not impose any limit on the U(1) heavy vector-boson $B'$ mass
  since this can easily be made heavy enough to satisfy constraints by introducing a new $f$ as discussed in Ref.~\cite{Gregoire:2003kr}.}
We study the points that satisfy all these criteria and show the character of these points as blue dots in Figs.~\ref{dots2HDM.FIG},~\ref{dotsMfMG.FIG},~\ref{dotskhttsbemal.FIG},~\ref{dotsffT.FIG}. 

Precision electroweak constraints on this model have been analyzed for example in Refs.~\cite{Gregoire:2003kr,Han:2005dz,Csaki:2003si}. 
To get some idea of the constraints that it may be imposing, we consider here the ``near-oblique'' limit discussed in Ref.~\cite{Gregoire:2003kr}, 
for which we impose the following additional requirement:
$M_{W'} > (1800~{\rm GeV})*(g_2^2-2 g^2)/(g_2^2-g^2)$ (from Eq.~(3.7) of Ref.~\cite{Gregoire:2003kr}).
We do not impose any constraints coming from $B'$ as this is more model-dependent as already pointed out. 
The points that satisfy these constraints (in addition to all the constraints above that the blue points satisfy)
are shown in Figs.~\ref{dots2HDM.FIG},~\ref{dotsMfMG.FIG},~\ref{dotskhttsbemal.FIG},~\ref{dotsffT.FIG} as green dots.
The LSS Lagrangian parameters and the resulting masses, couplings, and other quantities for 9 sample points among the green-dots that satisfy direct and precision electroweak constraints
are listed in App.~\ref{SampPts.APP}. 

In Fig.~\ref{dots2HDM.FIG} we show some 2HDM aspects. In the LSS model, for the points that satisfy the 
constraints, $\tan\beta$ is typically small, lying in the range $(0.3,5.4)$. 
The heavy scalar masses $m_A$, $m_H$ and $m_{H^\pm}$ become more and more degenerate as the $m_A$ scale increases.
This can be understood from Eq.~(\ref{mass-terms}), which gives $m_{A}^{2}-m_{H}^{2} = m_{h}^{2}$ and $m_{A}^{2} - m_{H^{\pm}}^{2} = \lambda_5^\prime v^2$,
and therefore $(m_A - m_H)$ falls smoothly like $1/m_A$ since $m_h$ is fixed to the experimentally measured value, and $(m_A - m_{H^\pm})$ has scatter due to its dependence on $\lambda_5$. 

In Fig.~\ref{dotsMfMG.FIG}, we show the heavy vector-boson and vector-like fermion sectors. 
As the blue dots show, $M_{t2}$ can be as light as around $750$~GeV which has good discovery prospects at the LHC. 
As shown by the green dots, imposing precision electroweak constraints raises the mass scale of the BSM states as expected. 
See App.~\ref{SampPts.APP} for some sample green-dot points.
Nevertheless, $M_{b'} = 948~$GeV, satisfies precision electroweak constraints, and LHC discovery is still possible.
We will demonstrate later that these vector-like fermions can induce significant $gg\phi$ effective couplings
at the 1-loop level.  
See Ref.~\cite{Gopalakrishna:2015wwa} for a more general analysis of this aspect.  
The U(1) heavy vector-boson $B'$ could be significantly lighter than the heavy SU(2) vector-boson $W'$, although its mass is quite model dependent 
as we have already commented.
A detailed discussion of the LHC signatures of the vector-like fermions and heavy vector-bosons in little-Higgs models are found for example in Refs.~\cite{Perelstein:2005ka,LHC-signature-constraints}.
The LHC signatures of the $t_2, t_3, b'$ will have many similarities with those studied for example in Refs.~\cite{Gopalakrishna:2011ef,Gopalakrishna:2013hua}, 
and the heavy vector bosons with the studies for example in Ref.~\cite{ourVecBoson.BIB}.

In Fig.~\ref{dotskhttsbemal.FIG} we show $\kappa_{htt}$ and $s_{\beta-\alpha}$, with the former defined below Eq.~(\ref{hYuk.EQ}) as the ratio of the $htt$ coupling to its SM value,
and the latter is the ratio of the $hVV$ couplings to the corresponding SM values as in Eq.~(\ref{CPeWW.EQ}), with $VV=\{W^+W^-,ZZ\}$.  
We see that the alignment limit discussed in the beginning of this section is satisfied very well. 
Curiously, for all points that satisfy the constraints, both $hff$ and $hVV$ are negative, 
i.e. opposite in sign to the SM.
As already explained, the LHC observables measured thus far are largely sensitive only to the relative sign of $hff$ and $hVV$ and therefore will allow flipping both signs.  
It will be important to find observables that are sensitive to flipping both $hff$ and $hVV$ signs;
this will be the subject of future work. 

For the points that satisfy experimental constraints, we determine the amount of fine-tuning in the model. 
One measure of fine-tuning is how sensitive $\hat v \equiv v/f$ is to the input model parameters.
Various measures of fine-tuning $f_T$ are possible, 
one of which is\footnote{This is along the lines defined for example in Ref.~\cite{Barbieri:1987fn}.} 
\beq
f_T^{-1} \equiv  {\rm Max}_{i}  \left\{ \left|  \frac{ \partial \log{\hat v^2}}{\partial \log{\alpha_i}} \right|  \right\}  \ ,
\label{fTdefn.EQ}
\eeq
where $\alpha_i$ are the 9 input parameters discussed above.
The $\hat v$ dependence on the input parameters can be obtained via Eq.~(\ref{fasv.EQ}) using Eqs.~(\ref{2HDMfLSS.EQ})~and~(\ref{tbeta.EQ}). 
In Fig.~\ref{dotsffT.FIG} we show the fine-tuning measure $f_T$ defined in Eq.~(\ref{fTdefn.EQ}) as a function of $f$. 
We find that all points that satisfy the constraints are fine-tuned at a level worse than about $2\, \% $,
and those that satisfy precision constraints worse than about $0.3\, \% $.
This is a surprisingly bad fine-tuning which we naively do not expect since $v^2/f^2 \sim 1/(16\pi^2)$ due to it being generated at 1-loop.
This can be seen from Eq.~(\ref{fasv.EQ}) which implies $\hat{v} \propto (\hat{b}^2 - \hat{m}_1^2 \tan\beta)$,
where $\hat{b}^2 \equiv b^2/f^2$ and $\hat{m}_1^2 \equiv m_1^2/f^2$,
with the right-hand-side being generated at 1-loop as seen explicitly in Eq.~(\ref{2HDMfLSS.EQ}). 
The reason it turns out to be so badly fine-tuned is because we find that for the points that satisfy the phenomenological constraints,
the $y_i$ and $g_2$ are large, overcoming the $1/(16\pi^2)$ suppression in $\hat b^2$ and $\hat m_{1,2}^2$ in Eq.~(\ref{2HDMfLSS.EQ}) and making them ${\cal O}(1)$.
Thus to get a small $v^2/f^2$, a cancellation between two ${\cal O}(1)$ quantities $\hat b^2$ and $(\hat m_{1}^2 \tan\beta)$ becomes necessary, fine-tuning the model.  
As expected, $f_T$ gets worse as $f$ increases.

We explore next the detection prospects of the new neutral scalar states $A$ and $H$ at the LHC. 
\begin{figure}
\centering
\includegraphics[width=0.45\textwidth]{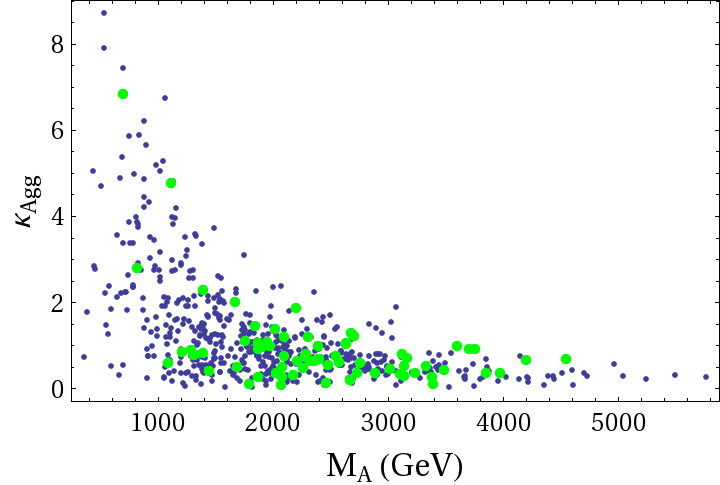}\hspace{1em}%
\includegraphics[width=0.45\textwidth]{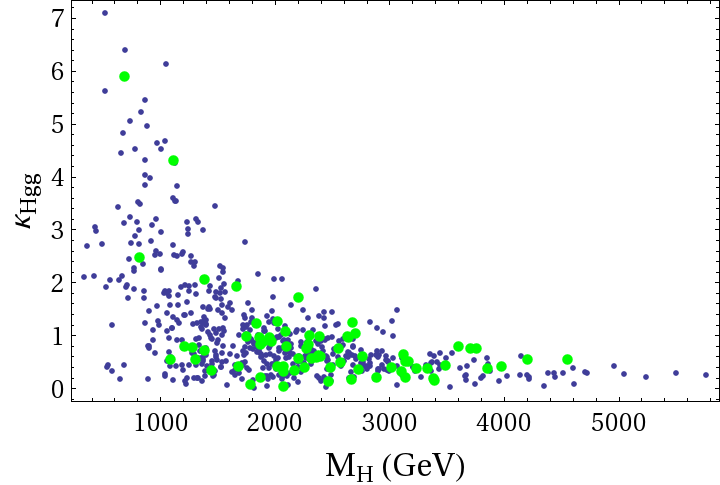}\hspace{1em}
\caption{$\kappa_{Agg}$ (left) and $\kappa_{Hgg}$ (right) for the allowed points of the parameter space. The color-coding of the dots is as in Fig.~\ref{dots2HDM.FIG}.}
\label{kphigg-mphi}
\end{figure}
To aid in this, in Fig.~\ref{kphigg-mphi}, we present the $\phi gg$ effective couplings, with $\phi = \{A, H\}$, in the notation defined in Ref.~\cite{Gopalakrishna:2015wwa}.
The 1-loop expressions for $\kappa_{\phi gg}$ and $\kappa_{\phi \gamma \gamma}$ are given in App.~\ref{phiggAA.APP}.
\begin{figure}
\centering
\includegraphics[width=0.45\textwidth]{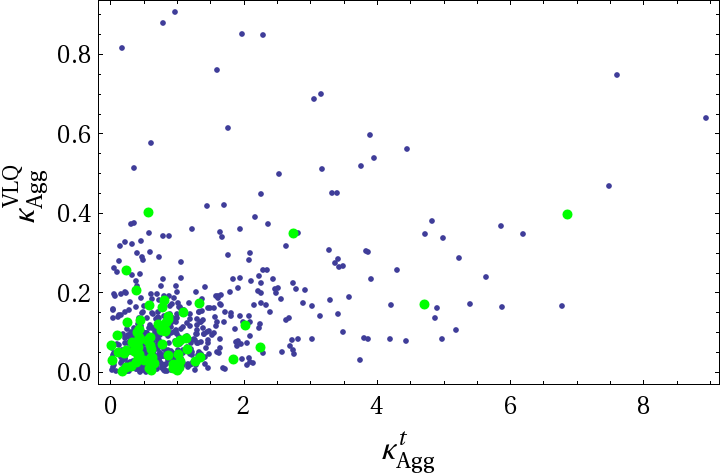}\hspace{1em}%
\includegraphics[width=0.45\textwidth]{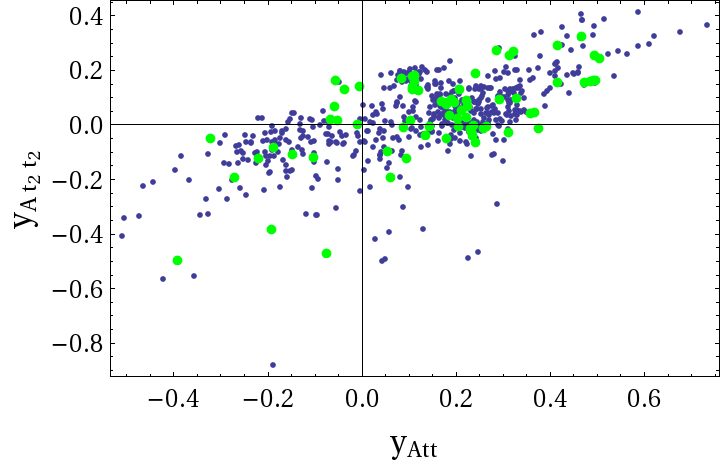}\hspace{1em}%
\caption{$\kappa_{Agg}^{t} $ vs. $ \kappa^{VLQ}_{Agg}$ (left), and $y_{Att}$ and $y_{A t_2 t_2}$ (right) for the allowed points of the parameter space.
The color-coding of the dots is as in Fig.~\ref{dots2HDM.FIG}.}
\label{kphiggVLFvSM.FIG}
\end{figure}
In Fig.~\ref{kphiggVLFvSM.FIG} (left) we present $|\kappa_{Agg}|$ with the top-quark (i.e. $t$) and VLQ (i.e. $ t_2$, $t_3$ and $b_2, b_3$) contributions separated. 
The top-quark contribution only is denoted by $\kappa_{Agg}^t$, and VLQ  contributions only denoted by $\kappa_{Agg}^{VLQ}$.
We separate these in this manner only for illustration purposes and the full amplitude is a coherent sum of these. 
We do not show the bottom-quark contribution since it is negligible due to $\tan\beta$ being not too large. 
We see that for some points the VLQ contributions can dominate over the top contribution.  
To support this further we show in Fig.~\ref{kphiggVLFvSM.FIG} (right) the couplings $y_{Att}$ versus $y_{At_2 t_2}$ in which we see that for some points the latter can be much larger
compared to the former.
These show that the vector-like fermion contributions to $Agg$ can be very important.   

We turn next to a discussion of the total width ($\Gamma$) and branching ratios (BR) of the scalars into SM final states.
Analytical expressions of all the relevant partial decay widths are given for example in Ref.~\cite{Gunion:1989we}.
The $BR(h\to XX)$ are all close to the SM case in the alignment limit and therefore satisfy the LHC constraints.
The $hbb$ coupling and BR could in principle be shifted, but for this coupling also to be SM-like in the alignment limit, it is sufficient for one of $\{y_b^{(1)}, y_b^{(2)}\}$ to be zero,
since in this case, the ratio of the $hbb$ coupling to the SM coupling is either $\sin\alpha/\cos\beta$ or its reciprocal (see Eq.~(\ref{LhHAbb.EQ})),
which are $\approx \pm 1$ in the alignment limit $(\beta-\alpha) = \pm \pi/2$ as discussed earlier.
In our analysis we assume that only $y_b^{(2)}$ is nonzero.
If both are nonzero, there will be a non-trivial constraint on the parameter-space coming from $BR(h\to bb)$, which we do not explore in this work.
Similar statements apply to the other (lighter) fermions ($\tau, c$). Thus, in this work, we explore the case when $y_{b,\tau,c}^{(1)} = 0$ and $y_{b,\tau,c}^{(2)}$ nonzero.

\begin{figure}
\centering
\includegraphics[width=0.32\textwidth]{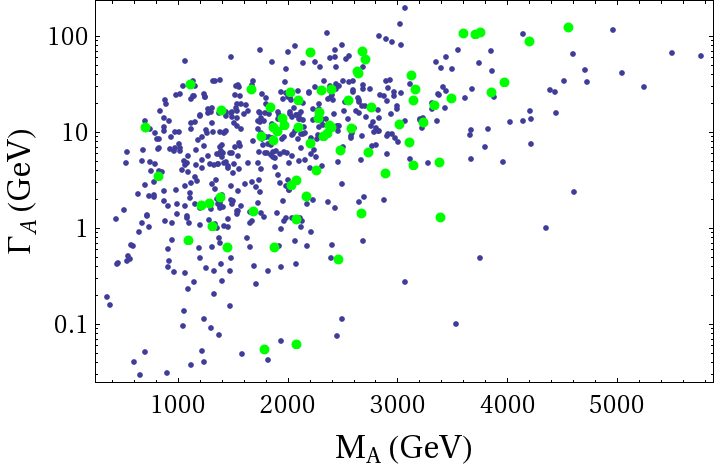}
\includegraphics[width=0.32\textwidth]{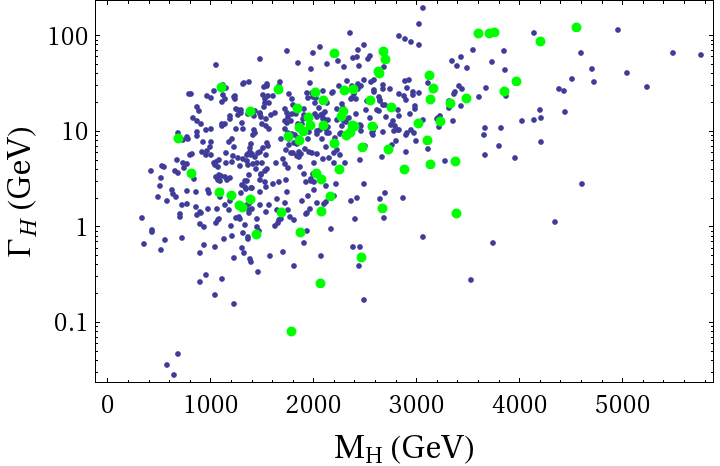}
\includegraphics[width=0.32\textwidth]{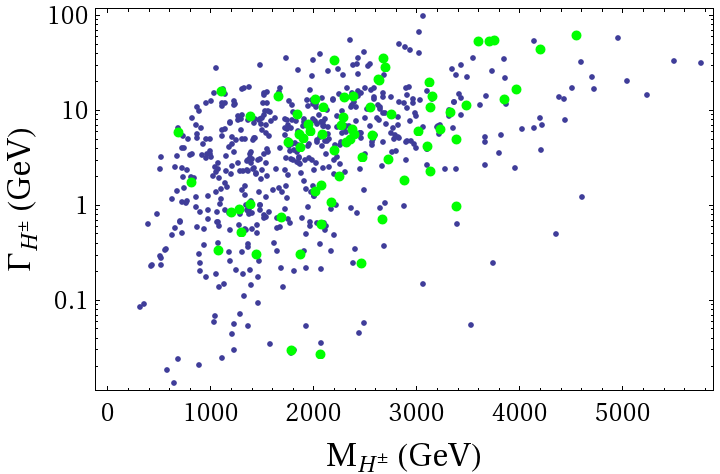}
\caption{$\Gamma_A$ (left), $\Gamma_H$ (middle), $\Gamma_{H^{\pm}}$ (right) for the allowed points of the parameter space.
  The color-coding of the dots is as in Fig.~\ref{dots2HDM.FIG}.
}
\label{AHHcwidth}
\end{figure}

In Fig.~\ref{AHHcwidth} we show $\Gamma_{A,H,H^\pm}$, the total widths of the heavy scalars, where
$\Gamma_A$ is a sum over the partial-widths to $tt,bb,cc,\tau\tau, Zh$ decay modes,
$\Gamma_H$ is over $tt,bb,cc,\tau\tau,WW,ZZ,hh$ decay modes, and
$\Gamma_{H^\pm}$ is over $tb,cs,\tau\nu,Wh$ decay modes.
Although generically $AVV$ is zero at tree-level while $HVV$ is not, $\Gamma_A$ and $\Gamma_H$ end up being almost identical.
This is because most of the allowed points satisfy the alignment limit to a very good degree, suppressing the tree-level $HVV$ couplings,
and at the same time the $H$ and $A$ couplings to the SM fermions become identical in this limit.

\begin{figure}
\centering
\includegraphics[width=0.45\textwidth]{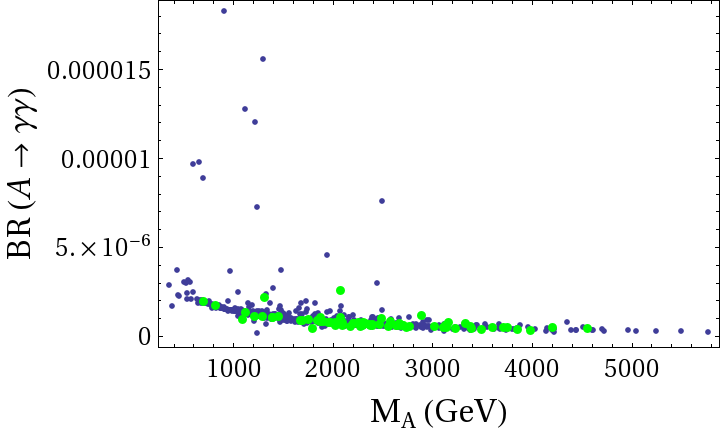}\hspace{1em}%
\includegraphics[width=0.4\textwidth]{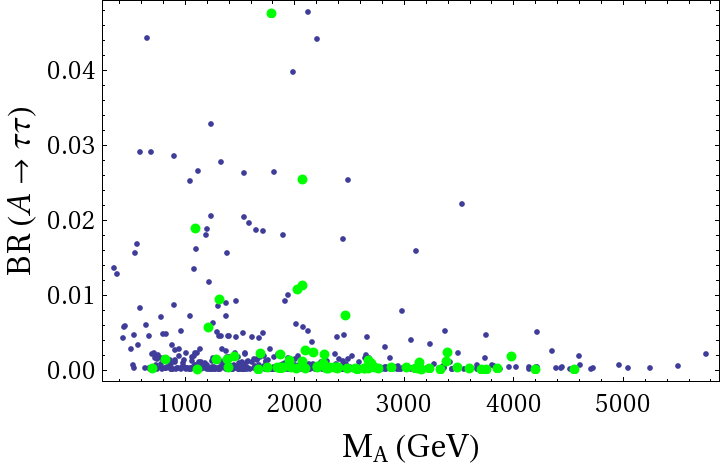}\hspace{1em}\\
\includegraphics[width=0.4\textwidth]{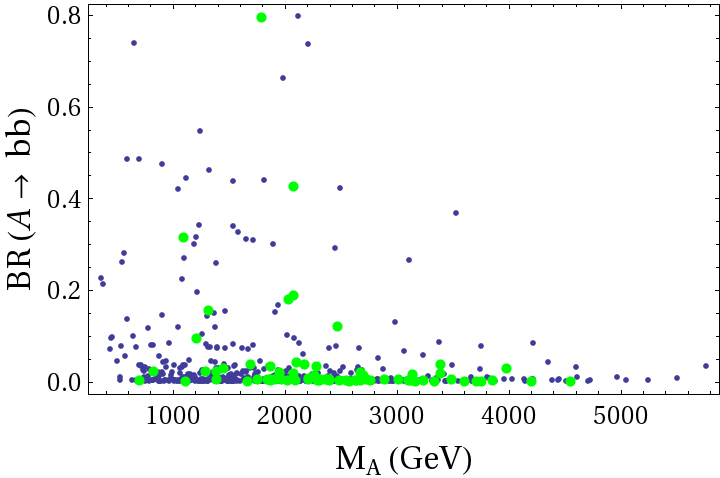}\hspace{1em}%
\includegraphics[width=0.4\textwidth]{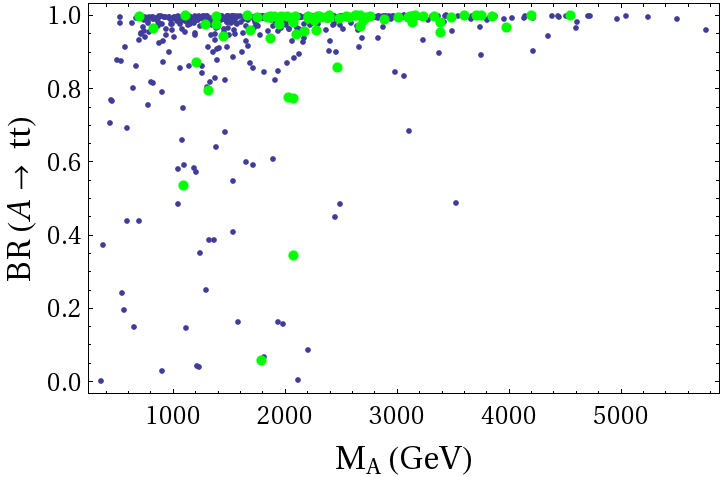}\hspace{1em}\\
\includegraphics[width=0.4\textwidth]{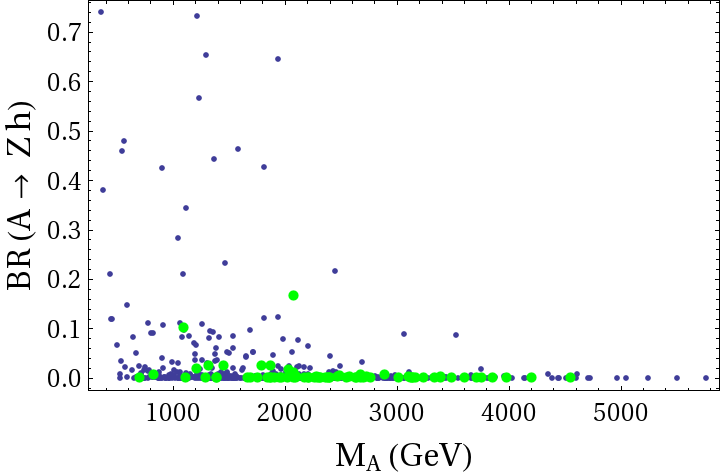}\hspace{1em}
\caption{BR~$(A\to \gamma \gamma)$ (top left), BR~$(A \to \tau \tau)$ (top right), BR~$(A\to bb)$ (middle left), BR~$(A \to tt)$ (middle right) and BR~$(A \to Z h)$ (bottom)
for the allowed points of the parameter space. The color-coding of the dots is as in Fig.~\ref{dots2HDM.FIG}.}
\label{brAlss}
\end{figure}

\begin{figure}
\centering
\includegraphics[width=0.45\textwidth]{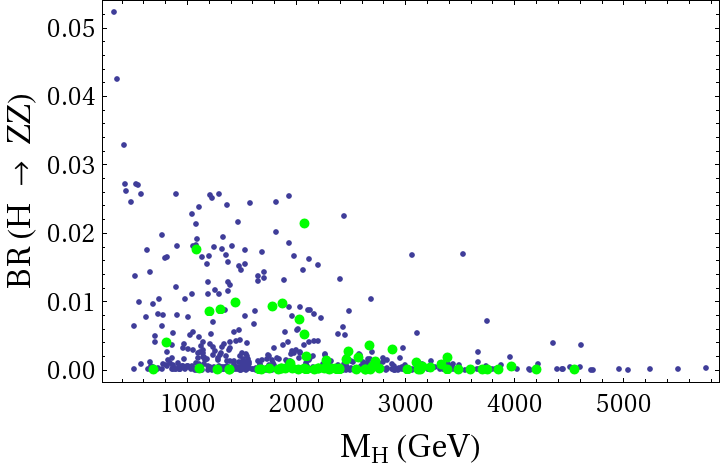}\hspace{1em}
\includegraphics[width=0.45\textwidth]{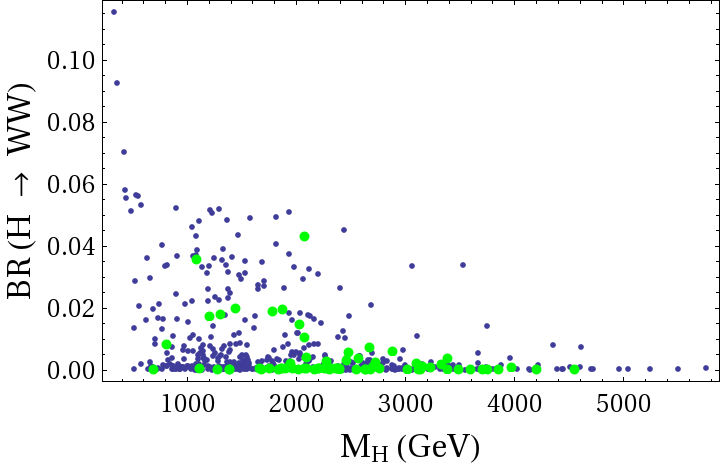}\\
\includegraphics[width=0.45\textwidth]{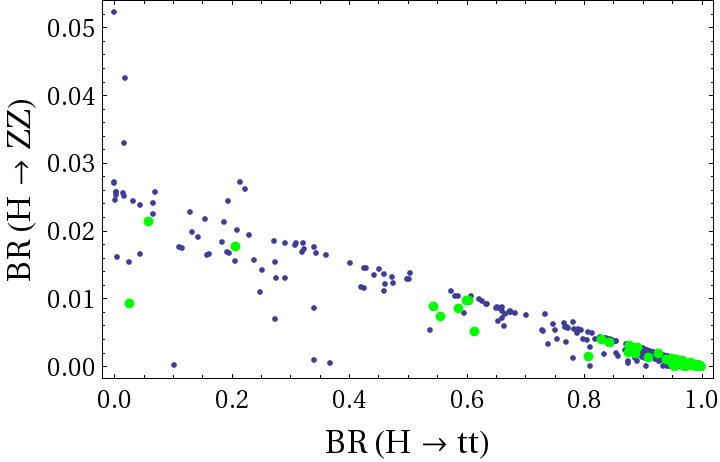}\hspace{1em}
\includegraphics[width=0.45\textwidth]{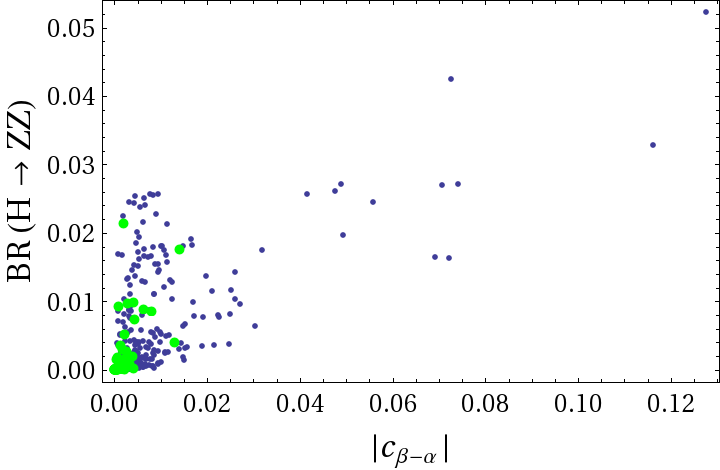}\\
\includegraphics[width=0.45\textwidth]{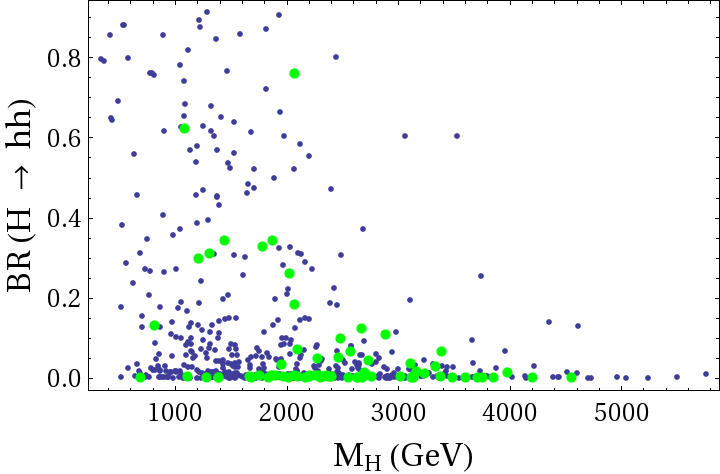}\hspace{1em}
\includegraphics[width=0.45\textwidth]{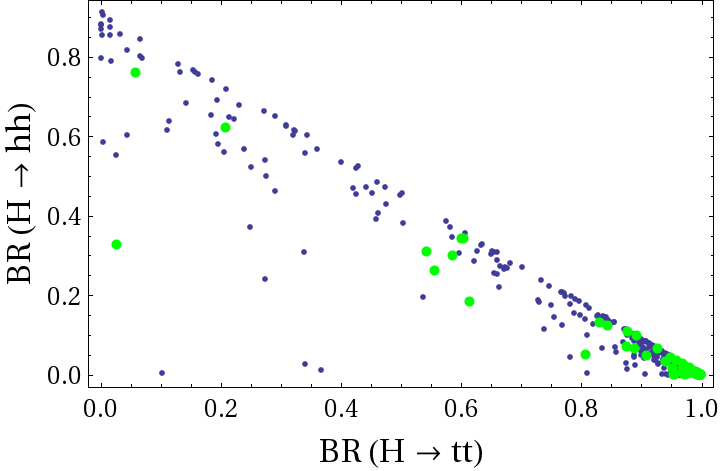}
\caption{ BR$(H \to ZZ)$ (top left), BR$(H \to WW)$ (top right), correlation between BR$(H \to tt)$ and BR$(H \to ZZ)$ (middle left),
correlation between BR$(H \to ZZ)$ and $|c_{\beta - \alpha}|$ (middle right),
BR$(H \to hh)$ (bottom left) and correlation of BR$(H \to hh)$ with BR$(H \to tt)$ (bottom right) for the allowed points of the parameter space.
The color-coding of the dots is as in Fig.~\ref{dots2HDM.FIG}.
}
\label{brHlss}
\end{figure}

In Fig.~\ref{brAlss} we show BR~$(A \to \gamma \gamma,\, \tau \tau,\, bb,\, tt,\, Z h)$ 
for the allowed points of the parameter space.
BR~$(A \to Z h)$ is very small in most part of the parameter space because
$AZh$ coupling is proportional to $c_{\beta - \alpha}$ which goes to zero in the alignment limit.
There are few points where the alignment limit is not perfect and at the same time $y_{Att}$ is small;
BR$(A \to Zh)$ becomes significant for these points.
We do not show explicitly the corresponding BR for the $H$ as they are quite similar to the $BR(A\to XX)$.
One important difference between the $A$ and $H$ is that at tree-level, the $AV V$ (with $V = {W, Z}$) couplings are zero
and are only generated by SM and BSM fermions at the loop level, while the $HV V$ couplings could be
nonzero at tree-level. However, in the alignment limit we consider the $HV V$ couplings
are zero. Thus in the alignment limit the $A$ and $H$ have very similar phenomenology.
One difference is in BR$(H \to \gamma \gamma$), for which $H^{\pm}$ loops also contribute.
However this contribution is small (see Ref.~\cite{Djouadi:2005gj}) and does not change the results qualitatively.
The largest value of $BR(H\to \gamma\gamma)$ is about $4.8 \times 10^{-6}$.
Away from the alignment limit $H$ can also decay to $WW,ZZ,hh$ at tree-level,
while BR$(H \to Zh)$ is identically zero.
In Fig.~\ref{brHlss} we plot BR$(H\to ZZ,WW,hh)$, and we see that these BRs can become sizable when either of the following two things happen:
(i) the alignment limit is not perfect making the $HZZ$ and $HWW$ significant, or, 
(ii) the $Htt$ coupling given below Eq.~(\ref{hYuk.EQ}) becomes accidentally small, in turn making BR$(H\to tt)$ small. 
To illustrate these effects, the correlation of BR$(H \to ZZ)$ and BR$(H \to hh)$ with $BR(H \to tt)$
and the correlation of BR$(H \to Z Z)$ with $|c_{\beta - \alpha}|$ is shown in Fig.~\ref{brHlss}; similar results hold for the $H \to W W$ channel.
If $m_{A,H} > m_{t_2,b_2} + m_{t,b}$ then $A,H \to t_2 t,~ b_2 b$ decays become kinematically allowed.
We do not analyze these modes in this paper as there is only one point (namely the second point of table~\ref{samplePts.TAB})
that satisfies $m_{A,H} > m_{b_2} + m_b$.

Having determined $\kappa_{\phi gg}$, we can use Figs.~1,~2,~3 in Ref.~\cite{Gopalakrishna:2015wwa} to know whether the point is allowed
by the $8~$TeV LHC run, and what the $A, H$ signal c.s. is at the $14~$TeV LHC.
For all the points in the LSS model that we found to satisfy the constraints, 
BR($\phi\to \gamma\gamma$) is too small to be interesting, $BR_{\tau\tau} \sim 10^{-2}$ makes this mode very challenging, 
and the $BR_{b\bar b}$ although reasonable, has a large QCD background.
This leaves the $t\bar t$ mode as a good possibility. 
For example, for the $m_{A,H}\approx 900~$GeV green dot in Fig.~\ref{kphigg-mphi}, $\kappa_{Agg,\, Hgg} \approx 2.5$, 
and from Ref.~\cite{Gopalakrishna:2015wwa}, we find that the LHC $8~$TeV constraints does indeed allow this point,
$\sigma(gg\to\phi) \approx 20~$fb at the $14~$TeV LHC, and
since $BR_{t\bar t}$ is sizable, the $t\bar t$ mode is the most promising one.  
A detailed analysis of the LHC signatures including backgrounds for some of the promising points in parameter space that we have identified here
will be the subject of a future study. 

\begin{figure}
\centering
\includegraphics[width=0.45\textwidth]{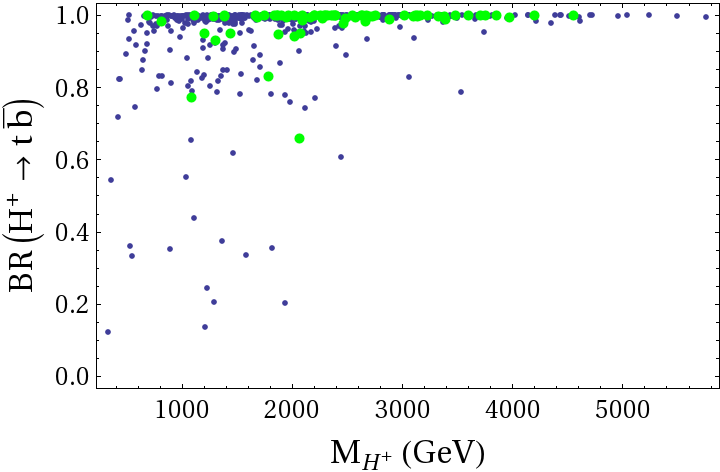}
\includegraphics[width=0.45\textwidth]{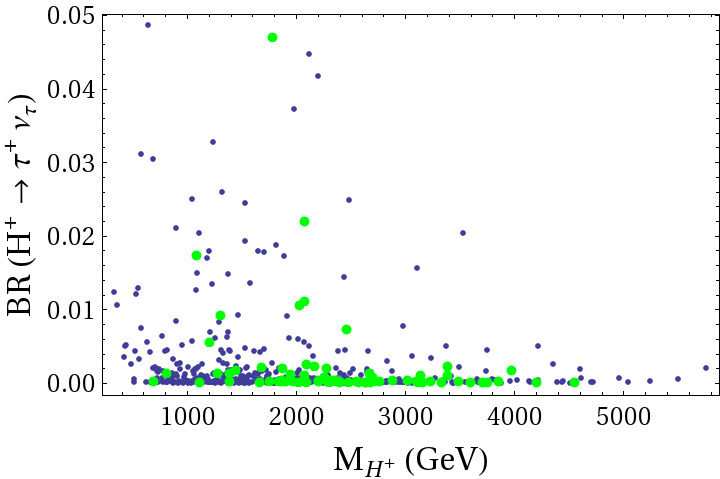}\\
\includegraphics[width=0.45\textwidth]{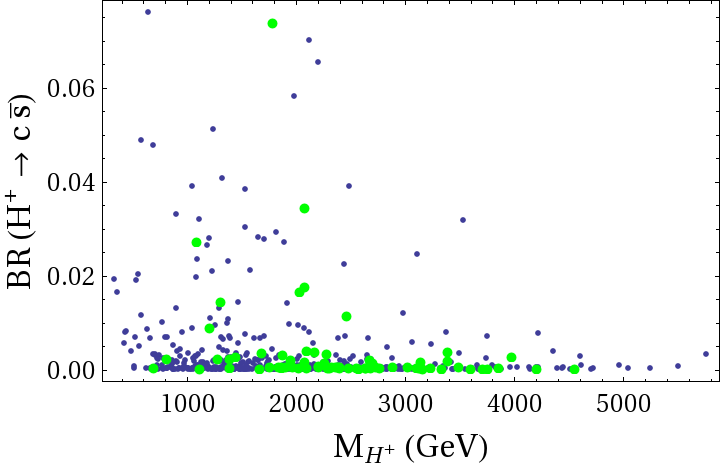}
\includegraphics[width=0.45\textwidth]{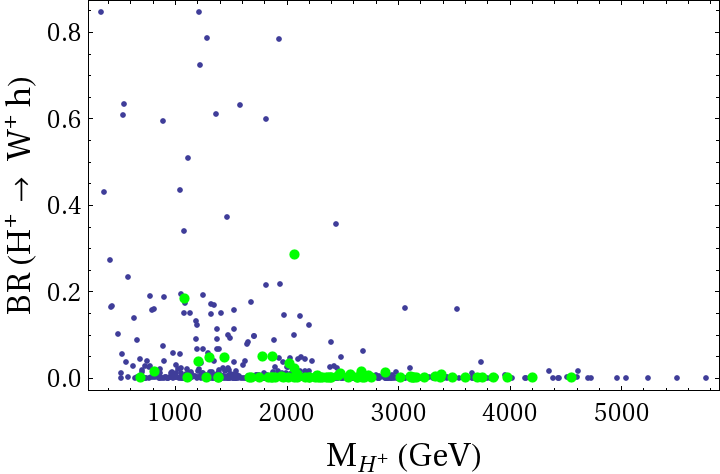}
\caption{ BR$(H^+ \to t \bar{b})$ (top left), BR$(H^+ \to \tau^+ \nu_{\tau})$ (top right), BR$(H^+ \to c \bar{s})$ (bottom left) and BR$(H^+ \to W^+ h)$ (bottom right) for the allowed points of the parameter space. The color-coding of the dots is as in Fig.~\ref{dots2HDM.FIG}.}
\label{brHclss}
\end{figure}

In Fig~\ref{brHclss} we show BR$(H^+ \to t \bar{b},~\tau^+ \nu_\tau,~c \bar{s},~W^+ h)$, assuming that only the $y^{(2)}$'s are nonzero.
BR$(H^+ \to t \bar{b})$ is the largest for most part of the parameter space since the $H^+ t b$ coupling is generically large.
For a few points however, when the $H^+ t b$ coupling becomes smaller due to partial cancellations between the various terms in $y^{+}_{00}$ of Eq.~(\ref{lHctb.EQ}), 
larger values of BR$(H^+ \to \tau^+ \nu_\tau, c \bar{s}, W^+ h
)$ are possible.
We expect BR$(H^+ \to c \bar{s})$ to be similar to BR$(H^+ \to \tau^+ \nu_\tau)$ since their coupling to $H^\pm$ is $(m_{c,\tau} \tan \beta/v)$; 
although the former is enhanced by a color factor of $3$, $m_c/m_\tau \approx 0.7$~\cite{Agashe:2014kda}, leading to BR$(H^+ \to c \bar{s}) \approx 3 * (0.7)^2 * BR(H^+ \to \tau^+ \nu_\tau)$.
The exclusion limit on $\sigma(H^+) \times$~BR$(H^+ \to \tau \nu_\tau)$ from Ref.~\cite{Aad:2014kga} does not put any further constraints on the parameter space of the LSS model.
The $H^{\pm} \to tb$ decay channel at CMS and ATLAS is discussed in Refs.~\cite{Hctb-lhc},
but there are no results yet for $m_{H^\pm} > 600~$GeV.
In the future, this can be a very promising channel of the $H^\pm$.
A detailed analysis of the LHC signatures of the $H^\pm$ are studied, for instance,
in Ref.~\cite{Basso:2012st} which studies it in the context of a CP-violating Type-II 2HDM, 
and
in Refs.~\cite{genYukCharged} which study its production and decay in a Type III 2HDM after including the $B \to X_s \gamma$ and perturbativity constraints.

Depending on the flavor structure of the Yukawa couplings of the other (lighter) fermions, flavor-changing-neutral currents (FCNC) can place important constraints on the model.
From Eq.~(\ref{Lferm.EQ}) it is clear that the top-quark couples to both $\phi_1$ and $\phi_2$, which implies a Type III 2HDM flavor structure, 
and we expect FCNCs involving the 3rd generation in particular to be the non-trivial ones.  
As we mentioned earlier, having $y_{b,\tau,c}^{(1)} \neq 0; y_{b,\tau,c}^{(2)} \neq 0$ will place non-trivial constraints from the $h\to bb,\tau\tau$ measurement at the LHC, 
and to avoid this, we considered the case when $y_{b,\tau,c}^{(1)} = 0; y_{b,\tau,c}^{(2)} \neq 0$. 
If this pattern is adopted for the other light fermions as well, we are in the framework of a Type I 2HDM for the light fermion sector,
with only the top breaking the Type I structure.
Similarly, we will be in the Type I framework for the light fermions if $y_{b,\tau,c}^{(1)} \neq 0; y_{b,\tau,c}^{(2)} = 0$.
Alternately, one could explore the case when only the $y^{(1)}\neq 0$ for the up-type light fermions, and only $y^{(2)}\neq 0$ for the down type
(and the top couples to both $\phi_1$ and $\phi_2$), which also satisfies $h\to bb,\tau\tau$ LHC constraints. 
In this case the light fermion sector will be analogous to a Type II 2HDM, with again only the top breaking the Type II structure.
An analysis of these flavor issues is beyond the scope of this work.
For a recent analysis of FCNCs in the Type III 2HDM (although at large $\tan\beta$) see for example Ref.~\cite{Crivellin:2013wna} and references therein.

There is very strong evidence for dark matter in astrophysical and cosmological observations. 
To have the possibility of dark matter, one can extend the model studied here to implement a discrete symmetry that results in a stable particle that can be identified as dark matter.
For instance, extending the $SU(6)/Sp(6)$ model to make it invariant under $T$-parity provides a stable dark matter candidate, while also softening constraints from precision electroweak observables, 
as has been studied in detail in Ref.~\cite{Brown:2010ke}. 
Some other extensions of little Higgs models that include dark matter are studied for example in 
Refs.~\cite{Hubisz:2004ft}-\cite{Cacciapaglia:2009cv}. 

\section{Conclusions}
\label{Concl.SEC}

Little-Higgs theories improve naturalness of the Higgs sector by removing the 1-loop quadratic divergence present in the Higgs sector of the standard model.
One example of a little-Higgs model is the $SU(6)/Sp(6)$ Low-Skiba-Smith (LSS) model, which we study in detail in this work.
Like in all little-Higgs models, new vector-like fermions ($t_2, t_3, b_2, b_3$) and heavy vector-boson states ($W',B'$) are present at around the TeV scale.
In addition, in the LSS model, the scalar sector is a 2-Higgs-doublet model (2HDM), which includes in the physical spectrum, two CP-even scalars ($h,H$), a CP-odd scalar ($A$) and a charged scalar ($H^\pm$). 
We identify the lighter CP-even scalar $h$ as the $125~$GeV state observed at the LHC. 
Its couplings to other SM states can be shifted, and therefore the 8~TeV LHC measurements place nontrivial constraints on the parameter space. 
We show in this paper that such constraints can be satisfied, and we present various properties of the heavy scalars that can be useful for searches at the 14~TeV LHC and future colliders. 

We begin with an effective 2HDM Lagrangian of the specific kind generated in the LSS model and identify the physical region of masses and couplings in the $m_A$--$\tan\beta$ plane.    
We then develop the relations of these 2HDM effective parameters in terms of the LSS Lagrangian parameters.
We perform a random scan of the LSS parameter space to identify points that satisfy direct collider constraints and precision electroweak constraints.
To adequately sample the 10-dimensional parameter space, we randomly sample the parameter space, and for each starting point, use a steepest descent algorithm to minimize a $\chi^2$ function given by the constraints.   
The direct 8~TeV LHC constraints we take into account are listed in Table~\ref{constrVals.TAB},
and include the Higgs mass, Higgs couplings to the top, bottom, $\tau$, $W^\pm$ and $Z$, top-quark mass, and LHC bounds on colored vector-like fermions ($t'$ and $b'$).
We present the Yukawa couplings that generate the $t,b,c,s$-quark and $\tau$-lepton masses since they decide the BR of the heavy scalars,
but do not specify the full flavor structure of the LSS model and do not work out the FCNC constraints that ensue.
We refer the reader to other works that have appeared recently investigating FCNCs in the 2HDM context.

For the points that satisfy the constraints, we present various aspects of the scalars, namely,
their masses, the deviations in the $hVV$ couplings, with $VV=\{W^+W^-,ZZ\}$, and how well the ``alignment limit'' (or decoupling limit) is satisfied,
the deviations in the $htt$ coupling, the fine-tuning required in the model, the 1-loop generated $\kappa_{Agg}, \kappa_{Hgg}$ couplings including SM and vector-like fermion contributions,
$BR(A \to \gamma\gamma, \tau\tau, bb, tt)$, $BR(H \to WW,ZZ,hh)$ and $BR(H^\pm \to tb,\tau\nu, cs, Wh)$.
We can use these $\kappa_{Agg,Hgg}$ in conjunction with the results of Ref.~\cite{Gopalakrishna:2015wwa} to ascertain whether these $A,H$ points are allowed by the $8~$TeV LHC run,
and obtain the $A,H$ production cross-section at the 14~TeV LHC. 
We find all points that satisfy these constraints to be fine-tuned worse that about $2\,\% $, and those that satisfy precision electroweak constraints in addition, to be worse than $0.3\,\% $.
$\tan\beta$ is roughly in the range $(0.3,5.4)$.
The alignment limit is found to hold to a very high degree in that the magnitude of the $hVV$ coupling is SM-like. 
Interestingly however, the $hVV$ coupling and $htt$ couplings are {\em both} flipped in sign compared to the SM, and since the LHC data is largely sensitive only to the relative sign, it allows this.
It will be important to find observables that are sensitive to the absolute sign of these couplings.

\medskip
\noindent {\it Acknowledgments:}
We thank V.~Ravindran for a discussion on QCD corrections to the top mass. 


\setcounter{section}{0}
\renewcommand\thesection{\Alph{section}}               
\renewcommand\thesubsection{\Alph{section}.\arabic{subsection}}
\renewcommand\thesubsubsection{\Alph{section}.\arabic{subsection}.\arabic{subsubsection}}

\renewcommand{\theequation}{\Alph{section}.\arabic{equation}}    
\renewcommand{\thetable}{\Alph{section}.\arabic{table}}          
\renewcommand{\thefigure}{\Alph{section}.\arabic{figure}}        

\section{1-loop $\kappa_{\phi gg}, \kappa_{\phi \gamma \gamma}$ effective couplings}
\label{phiggAA.APP}

The 1-loop expressions for the $\phi g g$ and $\phi \gamma\gamma$ amplitudes $\kappa_{\phi gg}$ and $\kappa_{\phi \gamma\gamma}$ respectively,
with $\phi = \{h,H,A\}$, as defined in Ref.~\cite{Gopalakrishna:2015wwa} are given here. 
These amplitudes are induced by quarks whose effective Lagrangian can be written as ${\cal L}^{f}_{\phi} \supset m_{f} \bar{f} f + y_{\phi ff} \phi \bar{f} f$.  
Defining $r_{f} = m_{f}^{2}/m^{2}_{\phi}$ and with $f$ running over all colored fermion species with mass $m_{f}$ and Yukawa couplings $y_{\phi ff}$, 
and with the electric charge of the fermion ($f$) denoted by $Q_{f}$,
the general expressions for $\kappa_{\phi gg}$ and $\kappa_{\phi \gamma\gamma}$ are given as
\bea
\label{kphiag}
\kappa_{\phi \gamma \gamma} = 2 e^2 \sum_{f} N_{c}^f Q_{f}^{2} \, y_{\phi ff} \frac{M}{m_f}  F_{1/2}^{(1)}(r_{f}) \ , \qquad
\kappa_{\phi gg} =  g_{s}^{2} \sum_{f} y_{\phi ff} \frac{M}{m_f}  F_{1/2}^{(1)}(r_{f})  \ , \\
{\rm\text with} \ \ 
F_{1/2}^{(1)}(r_{f}) = 4 r_{f} \left(\int_{0}^{1}dy \int_{0}^{1-y} dx \frac{g(x,y)}{(r_{f}-x y)}\right) \ , \nonumber
\eea
with $g(x,y)= (1 - 4 x y)$ for the CP-even scalars ($h,H$) and $1$ for the CP-odd scalar ($A$)
and $M$ is a mass scale which we set to $1~$TeV for numerical results.
We have used these expressions for the LSS model discussed in the text. 

\section{Sample points}
\label{SampPts.APP}

Here we present some sample points that satisfy direct collider and precision electroweak constraints discussed in Sec.~\ref{LSSpheno.SEC}.
Of the 70 points that went into the plots of Sec.~\ref{LSSpheno.SEC} as green-dots, we present 9 points here. 
\begin{table}[ht]
  \caption{Some sample points that satisfy direct collider and precision electroweak constraints.
    The corresponding quantities for the points shown in the upper table is continued in the lower table. 
    The $f$ and all masses are in GeV. 
  \label{samplePts.TAB}}
\begin{centering}
%
{\small
\begin{tabular}{|c||c||c|c|c|c|c|c|c|c|c||c|c|c|c|}
\hline 
No. & $f$ & $g_{1}$ & $g_{1}^{\prime}$ & $y_{1}$ & $y_{2}$ & $y_{3}$ & $y_{4}$ & $y_{5}$ & $c$ & $c'$ & $g_{2}$ & $g_{2}^{\prime}$ & $t_{\beta}$ & $\lambda_{5}^{\prime}$\tabularnewline
\hline 
\hline 
1 & 633.5 & 0.657 & 0.567 & 1.985 & 1.342 & 2.372 & 0.165 & 2.076 & 1.33 & 1.771 & 6.864 & 0.465 & 0.73 & 0.568\tabularnewline
\hline 
2 & 799.8 & 0.655 & 2.869 & 2.774 & 1.422 & 1.946 & 0.358 & 1.185 & 1.325 & 2.579 & 10.9 & 0.363 & 0.73 & 0.566\tabularnewline
\hline 
3 & 1096. & 0.688 & 1.6 & 1.664 & 1.199 & 2.691 & 0.843 & 2.706 & 1.51 & 2.246 & 2.106 & 0.369 & 0.6 & 0.666\tabularnewline
\hline 
4 & 1133. & 0.901 & 2.809 & 2.276 & 1.343 & 2.082 & 0.533 & 1.977 & 0.775 & 2.46 & 0.949 & 0.363 & 0.74 & 0.561\tabularnewline
\hline 
5 & 1133. & 0.851 & 1.448 & 1.568 & 1.257 & 2.391 & 0.798 & 2.602 & 0.805 & 2.172 & 1.021 & 0.371 & 0.95 & 0.513\tabularnewline
\hline 
6 & 1161. & 0.675 & 1.23 & 2.305 & 1.475 & 2.027 & 0.485 & 1.887 & 1.169 & 1.596 & 2.634 & 0.376 & 0.88 & 0.509\tabularnewline
\hline 
7 & 1207. & 0.852 & 1.449 & 1.583 & 1.271 & 2.393 & 0.800 & 2.612 & 0.808 & 2.173 & 1.018 & 0.371 & 0.97 & 0.517\tabularnewline
\hline 
8 & 1403. & 0.679 & 1.161 & 2.763 & 2.525 & 2.113 & 1.64 & 1.464 & 1.587 & 2.681 & 2.395 & 0.378 & 1.79 & 0.713\tabularnewline
\hline 
9 & 1429. & 0.705 & 1.376 & 2.607 & 2.82 & 2.234 & 2.048 & 1.507 & 1.558 & 2.504 & 1.742 & 0.373 & 1.9 & 0.751\tabularnewline
\hline 
\end{tabular}
\vspace*{0.25cm}
\begin{tabular}{|c||c|c|c|c|c|c|c|c|c|c|c|c|c|c|}
\hline 
No. & $M_{s}$ & $M_{W'}$ & $M_{B'}$ & $m_{h}$ & $m_{H}$ & $m_{A}$ & $m_{H^{\pm}}$ & $m_{t}$ & $M_{t2}$ & $M_{t3}$ & $M_{b2}$ & $M_{b3}$ & $\kappa_{htt}$ & $s_{\beta-\alpha}$\tabularnewline
\hline 
\hline 
1 & 5163. & 3089. & 328.6 & 124.8 & 1111. & 1118. & 1111. & 166.4 & 1218. & 1794. & 1315. & 1727. & -1.007 & -1. \tabularnewline
\hline 
2 & 10220. & 6177. & 1636. & 124.7 & 1666. & 1671. & 1666. & 159.2 & 2376. & 1823. & 947.5 & 1928. & -0.987 & -1. \tabularnewline
\hline 
3 & 3573. & 1716. & 1272. & 124.7 & 1445. & 1450. & 1443. & 161.1 & 2037. & 3246. & 2965. & 3228. & -1.011 & -1. \tabularnewline
\hline 
4 & 2719. & 1048. & 2269. & 124.5 & 1901. & 1905. & 1901. & 159.1 & 2537. & 2935. & 2239. & 2806. & -0.98 & -1. \tabularnewline
\hline 
5 & 2496. & 1065. & 1198. & 124.4 & 1284. & 1290. & 1284. & 158.4 & 1987. & 3078. & 2949. & 3061. & -0.969 & -1. \tabularnewline
\hline 
6 & 4039. & 2231. & 1056. & 123.1 & 2020. & 2024. & 2020. & 159.7 & 2626. & 3036. & 2190. & 2910. & -0.977 & -1. \tabularnewline
\hline 
7 & 2681. & 1133. & 1276. & 125. & 1386. & 1392. & 1386. & 159.4 & 2134. & 3287. & 3152. & 3271. & -0.974 & -1. \tabularnewline
\hline 
8 & 7283. & 2470. & 1212. & 125. & 2075. & 2079. & 2074. & 163.3 & 4402. & 4748. & 2055. & 4620. & -0.998 & -1. \tabularnewline
\hline 
9 & 7202. & 1898. & 1440. & 124.1 & 1309. & 1315. & 1307. & 162.7 & 4707. & 5191. & 2153. & 5141. & -1.011 & -1.\tabularnewline
\hline 
\end{tabular}
} 
\par\end{centering}
\end{table}


\end{document}